\documentclass[journal,onecolumn]{IEEEtran}
\usepackage{graphics}
\usepackage{amssymb}
\usepackage{graphicx}
\usepackage{fancybox}
\usepackage{amsmath}
\usepackage{subfigure}
\usepackage{algorithm}
\usepackage{algorithmic}
\usepackage{cite}
\usepackage{arev}
\usepackage{lmodern}
\usepackage{multirow}
\usepackage{booktabs}
\usepackage{etoolbox}
\usepackage{tabularx}
\usepackage{bm}
\usepackage{hyperref}

\usepackage{subcaption}
\usepackage{setspace}

\usepackage[mode=buildnew]{standalone}
\usepackage{xcolor}
\usepackage{tikz}
\usetikzlibrary{dsp}
\usetikzlibrary{chains}
\usepackage{pgfplots}
\newcommand{\VEC}[1]{\mathbf{#1}}          
\newcommand{\VECG}[1]{\boldsymbol{#1}}     
\newcommand{\MAT}[1]{\mathbf{#1}}          
\newcommand{\MATG}[1]{\boldsymbol{#1}}     

\newcommand{\putindex}[3]{\vtop{\hbox{\hspace{#3} $#1$}
            \hbox{\raise 6mm \hbox{$\scriptscriptstyle #2$}}}}

\newcommand{\gradx}[0]{\vtop{\hbox{\rm grad}
            \hbox{\raise 2.5mm \hbox{\rm \hspace{2mm} \footnotesize x}}}}

\newcommand{\grady}[0]{\vtop{\hbox{\rm grad}
            \hbox{\raise 2.5mm \hbox{\rm \hspace{2mm} \footnotesize y}}}}

\newcommand{\grad}[1]{\vtop{\hbox{\rm grad}
            \hbox{\raise 2.5mm \hbox{#1}}}}

\newcommand{\C}{\mathbb{C}}

\newcommand{\EV}[1]{E\{{#1}\}}                    

\newcommand{\btb}{     \begin{tabbing}             }
\newcommand{\bte}{     \end{tabbing}               }

\usetikzlibrary{shapes.geometric}
\usetikzlibrary{calc}

\definecolor{lightgreen}{RGB}{153,255,140}
\definecolor{lightblue}{RGB}{145,212,240}

\usepackage{multicol}

\usepackage{fancyhdr}

\author{Ernst Seidel$^{\ast}$, Gerald Enzner$^{\circ}$, Pejman Mowlaee$^{\vardiamond}$, Tim Fingscheidt$^{\ast}$ \\
{\small e.seidel@tu-bs.de, gerald.enzner@uni-oldenburg.de, pmowlaee@jabra.com, t.fingscheidt@tu-bs.de} \\[0.3cm]

$^{\ast}$Institute for Communications Technology,
Technische Universität Braunschweig,\\
Schleinitzstraße 22,
38106 Braunschweig, Germany\\ 
$^{\circ}$Department of Medical Physics and Acoustics, Division of Speech Technology and Hearing Aids, \\ Carl von Ossietzky Universität Oldenburg, 
 26111 Oldenburg, Germany\\
$^{\vardiamond}$GN Advanced Science,
Lautrupbjerg 7,
2750 Ballerup, Denmark
}

\newcommand\copyrighttext{%
  \footnotesize \textcopyright 2024 IEEE.  Personal use of this material is permitted.  Permission from IEEE must be obtained for all other uses, in any current or future media, including reprinting/republishing this material for advertising or promotional purposes, creating new collective works, for resale or redistribution to servers or lists, or reuse of any copyrighted component of this work in other works.}
\newcommand\copyrightnotice{%
\begin{tikzpicture}[remember picture,overlay]
\node[anchor=south,yshift=10pt] at (current page.south) {\fbox{\parbox{\dimexpr\textwidth-\fboxsep-\fboxrule\relax}{\copyrighttext}}};
\end{tikzpicture}%
}

\begin{document}
\title{\!Neural Kalman Filters for Acoustic Echo Cancellation\!

}

\maketitle
\copyrightnotice

\doublespacing

\begin{abstract}
Kalman filtering is a powerful approach to adaptive filtering for various problems in signal processing. The frequency-domain adaptive Kalman filter (FDKF) based on the concept of the acoustic state space provides a unifying solution to the adaptive filter update and the related stepsize control. It was conceived for the problem of acoustic echo cancellation and as such is frequently applied in hands-free systems. This article motivates and briefly recapitulates the linear FDKF and investigates how it can be further supported by deep neural networks (DNNs) in various ways, specifically, to overcome the challenges and limitations related to the usually required estimation of process and observation noise covariances for the Kalman filter.
While the mere FDKF comes with very low computational complexity, its neural Kalman filter variants may deliver faster (re)convergence, better echo cancellation, and even exceed the FDKF in its excellent double-talk near-end speech preservation both under linear and nonlinear loudspeaker conditions. To provide a synopsis of the state of the art, this article contributes a comparison of a range of DNN-based extensions of FDKF in the same training framework and using the same data.

\end{abstract}

\section{Introduction}
\label{sec:Intro}

Adaptive filters are widely used in signal processing to serve the tasks of signal estimation or system identification. In this article, we focus on adaptive filters operating on a (non)linear system output mixed with interferences, but also having access to the system input signal as reference. Prominent tasks of this configuration are system identification and echo removal, whereby the echo represents the (non)linear system output. 

\begin{figure}[t]
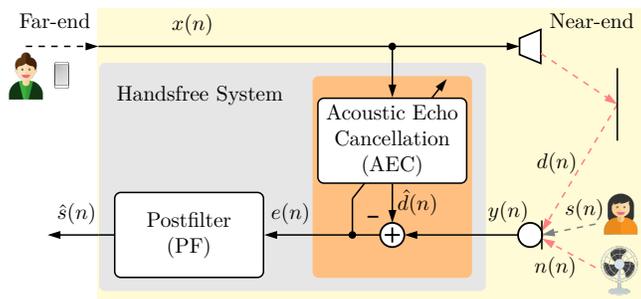

    \centering
    \includestandalone[width=0.5\textwidth, mode=image|tex]{fig/fig1}
    \vspace{-0.2mm}
	\caption{Generalized overview of a hands-free system / speakerphone. The speech signal of the far-end speaker is played out of a loudspeaker at the near-end, picked up at the microphone alongside near-end speech and background noise. An acoustic echo canceller and a postfilter (PF) aim at removing the echo and the background noise.}
	\label{fig:systemlvl}
 \end{figure}

One example application of such reference-based adaptive filters is the single-channel hands-free problem as shown in Fig.\ \ref{fig:systemlvl}. 
Here, a far-end (FE) speaker's voice, i.e., the reference signal $x(n)$, is presented by a (non)linear loudspeaker and picked up as echo by a microphone along with background noise and a near-end (NE) speaker's voice. To prevent the far-end speaker from echoing his/her own voice, typically an acoustic echo cancellation (AEC) algorithm is employed to estimate and to remove the disturbing echo from the microphone signal as part of a hands-free system / speakerphone. Various approaches to adaptive filters for AEC are presented in the following.

\subsection{Model-Based Approaches to System Identification and Acoustic Echo Cancellation (AEC)}
\label{ssec:Problem}

As depicted in Fig.\ \ref{fig:systemlvl}, for the task of AEC, we consider the microphone signal $y(n) = s(n) + n(n) + d(n)$ consisting of the target near-end speech signal $s(n)$, the background noise signal $n(n)$, and the echo signal $d(n)$ from the far-end speech signal $x(n)$, where $n$ is the discrete-time index. AEC aims at echo estimation and its subtraction from the microphone signal with an optional postfilter (PF) for residual echo and/or noise suppression. Estimating the echo $d(n)$ is---especially for classical approaches---often achieved through linear system identification of the (time-variant) room impulse response (RIR) $\VEC{h}(n)=\left( h_0(n), h_1(n), \dots, h_{N\!-\!1}(n) \right)^T$ of length $N$, which describes the echo generation from the vector of past reference samples $\VEC{x}(n)=\left( x(n), x(n\!-\!1), \dots, x(n\!-\!N\!+\!1) \right)^T$ in the near-end acoustic system through $d(n) = \VEC{x}^T(n) \VEC{h}(n)$. We denote the transpose by $(\cdot)^T$ and vectors/matrices by {bold symbols}.

Historically, adaptive filters for AEC emerged as time-domain normalized least mean-square (NLMS) \cite{Haensler2004}, multirate/subband \cite{Kellermann88,Breining99}, 
recursive least-squares (RLS), 
frequency-domain adaptive filtering (FDAF), 
 and multi-delay filtering (MDF) techniques  \cite{Benesty2001}. 
Coherently, these adaptive systems face the difficulty of attributing a larger error signal power to either RIR changes (thus requiring a larger stepsize for accelerated filter adaptation) or to near-end speech activity (then requiring a smaller stepsize for slower adaptation). 
It was the frequency-domain adaptive Kalman filter (FDKF)~\cite{Enzner2006} to consolidate for the first time the traditional adaptive filter process and the additional means for adaptive stepsize control under one umbrella, which allowed for a model-based optimal balance of tracking ability for the RIR and robustness against double-talk.

\subsection{From the Model-Based to the Data-Driven Paradigm---and to their Union}

With the advent of deep neural networks (DNN), also the field of acoustic echo control has been overwhelmingly captured 
by data-driven methods \cite{Richard2023}. 
Various DNN-based architectures were described to support or to replace model-based processing paradigms. 
DNN-based solutions were deployed to replace the echo estimate of the classical echo canceller \cite{Franzen2021, Seidel2021, Braun2022}, to provide a (spectral) mask for the microphone signal \cite{Seidel2023},
or to directly map to an echo-reduced speech output \cite{Wang2019_entryGE,Franzen2021}.
Some works even propose a single DNN for both echo and noise suppression jointly. 
Hybrid approaches employ a linear AEC followed by a DNN-based postfilter \cite{Valin_AEC,halimeh_combining_2021,ivry_deep_2021}.

In contrast, this article focuses on classical AEC, specifically the model-based FDKF approach, \textit{supported} by DNNs in various roles, which we refer to as \textit{neural Kalman filters.} The references will be provided on pp.\ \pageref{sec:Hybrid}ff., while our AEC simulation framework for comparison will be described on pp.\ \pageref{sec:Comparison}ff. We note here that comprehensive parallel developments of neural Kalman filters also exist outside the AEC context for  accomplishing enhanced state estimation under nonlinear dynamics and with partial information of the underlying state-space model \cite{Revach2022}. Moreover, there are approaches beyond neural Kalman filters to entirely estimate RIR filter coefficients by a DNN, such as \texttt{Meta-AF}~\cite{MetaAF2023}, where a self-supervised learning model replaces conventional adaptive filter updates.

The new trend of data-driven neural networks 
has somewhat disrupted the former and seemingly opposite trend of deliberate model-based processing for acoustic echo cancellation. The model here refers to the characterization of the acoustic enclosure of a hands-free system as an acoustic state space.
Owing to the availability of rich theoretical background, the state-space framework provides considerable support for rigorous treatment of time-varying acoustic systems in the presence of observation noise. Note that the time-variant aspect of hands-free systems refers to the important problem of acoustic echo path changes, while the observation noise may typically comprise the phenomena of ambient noise and double talk. The framework of acoustic state-space modeling unites both of these issues and therefore paves the way for an optimal balance of echo path tracking and robustness against observation noise. The concept of acoustic state-space modeling and the related derivation of pioneering algorithms, such as the FDKF, traces back to the research reported in \cite{Enzner2006,Enzner2010} and was followed up by model-based extensions \cite{Paleologu2013,Jung2014, yang2017frequency,Schrammen2019} until the advent of neural networks in this field.
Surprisingly, the pendulum now swings to a fascinating interplay of data-driven neural networks and model-based Kalman filters, i.e., the architectures of so-called \textit{neural Kalman filters} as shown by this article. Note the different motivations of the model-based and the data-driven paradigm, i.e., we have the physically-inspired state space on the one hand, and 
deep learning---encouraged by technological advancements---on the other hand. 
{Both motivations may complement each other very well for the design of \textit{neural Kalman filters}.}

\subsection{Real-World Challenges and the Trend Towards DNN-Supported (Neural) Kalman Filters}
\label{ssec:RealWorld}

AEC remains a challenging application of acoustic system identification, while---due to loudspeaker nonlinearities---having only weak access to the \textit{acoustic} system input. Moreover, to this day, it must cope with highly time-varying RIRs, nonpersistent system excitation, and near-end speech plus ambient noise interference. 
Acoustic Echo Cancellation Challenges have been organized at INTERSPEECH and ICASSP since 2021 \cite{Cutler2023}, providing comprehensive data and formal rules to foster and structure innovations in the field of AEC and more recently to achieve reproducible real-time methods evaluated on real-world data. The most successful methods perform mask-based echo suppression or residual echo suppression, while Kalman filters (KFs) alone would not have appeared top-ranked due to limitations when it comes to system nonlinearities or to the treatment of larger room sizes (i.e., longer RIRs).

On the other hand, the FDKF can deliver the target near-end speech component of the microphone signals almost distortionless at its output \cite{Franzen2018a,Seidel2024}. Moreover, the FDKF is of extremely low complexity and parameter footprint compared to DNN-based methods.
Another advantage of model-based processing for AEC (by its construction from prior statistical assumptions) might be the better generalization to unseen data.
      
For the many amenities of model-based processing, we observe a tremendous renaissance 
of Kalman filters for AEC, but \textit{paired with DNN support}, specifically to relieve major KF limitations consisting in the model linearity and in the difficulty of obtaining an a-priori acoustic state-space representation (i.e., the model covariances) for the practical Kalman gain computation \cite{yang2017frequency}. 
These kinds of hybrid approaches of model-based processing and deep learning are still of moderate complexity and memory footprint, delivering an excellent target near-end speech component and keeping their capability for system identification. This article outlines the range of efforts made on integrating DNNs into Kalman filters for AEC, reaching to a unified representation for depicting their similarities and differences and to discussions on their potential to push their current limits.

\section{Acoustic State Space and Related Kalman Filters for AEC}
\label{sec:Kalman}

We revisit the idea of adaptive filtering as a baseline, then introduce the concept of the acoustic state space for simultaneous representation of time-varying echo paths and double-talk interference, and finally depict the related Kalman filter algorithm, specifically FDKF, as a current workhorse for acoustic echo cancellation.

\subsection{Linear Signal Model and Basic Adaptive Filter}

Consider a linear time-varying system between the loudspeaker signal $x(n)$ and the observed microphone
signal $y(n)$ in Fig.~\ref{fig:systemlvl}. This configuration
with known input-output pairs of signals poses a supervised acoustic system identification problem. 
With the convolution index $\nu$ we have the microphone signal
\begin{align}
y(n)    & \,\, =\,\,   s(n) + n(n) + \sum\nolimits_{\nu=0}^{N-1} h_\nu(n)x(n-\nu) 
        \,\, =\,\,   s(n) + n(n) + \VEC{x}^T(n)\,\VEC{h}(n)  \;,
\label{measurements}
\end{align}
where the latent state $\VEC{h}(n)$ is subject to estimation from the available system output 
$\VEC{y}(n) = (y(n),y(n\!-\!1),\dots,y(0)) $
up to and including the current time $n$. For sequential RIR tracking at time $n$, the error signal
\begin{equation}
    e(n)= y(n) - \VEC{x}^T(n)\, \widehat{\VEC{h}}(n\!-\!1)
    \label{eq:error_signal}
\end{equation}
based on a former RIR estimate $\widehat{\VEC{h}}(n\!-\!1)$ may be considered. Gradient descent based on $\VECG{\nabla}_h e^2(n) \,\,=\,\, \partial e^2(n) / \partial \VEC{h}(n)$ has been a widely used baseline and with (\ref{eq:error_signal}) amounts to the celebrated least mean-square (LMS) algorithm
\cite{Haykin2002}
\begin{align}
    \widehat{\VEC{h}}(n) \,\, =\,\, \widehat{\VEC{h}}(n-1) - \frac{1}{2}\mu\VECG{\nabla}_h e^2 (n)
                       \,\, =\,\, \widehat{\VEC{h}}(n-1) + \mu e(n)\VEC{x}(n) \;,
\label{lms}
\end{align}
where $\mu$ is a relatively small non-negative stepsize factor. Yet more popular than the LMS is the normalized least mean-square (NLMS) algorithm, where the adaptive stepsize $\mu = \mu_0 /
||\VEC{x}(n)||^{2}$ levels out an input-signal dependent convergence rate of LMS. In this case, we have choices of $0<\mu_0<2$ for stability and $\mu_0=1$ for the highest rate of convergence under the conditions and assumptions detailed in  \cite{Haykin2002}.

\subsection{Acoustic State Space and Related Kalman Filter}

Finding an adaptive stepsize factor $\mu_0$ to support robustness during double-talk (by reducing $\mu_0$) and rapid convergence during echo path change (by increasing $\mu_0$) has occupied the AEC research community for decades, as shown, for instance, by the textbooks \cite{Gay2000,Haensler2004,Benesty2001,Vary2006}. In almost all cases the stepsize was determined from the available signals $x(n)$ and $y(n)$ and supplied as external input to the update equation of LMS, NLMS, RLS, FDAF, or the like. This treatment rests upon the limitation of the observation model in \eqref{measurements} to refer to any domain-specific quantification of the echo-path variability and the observation noise power. To achieve a balanced treatment of both phenomena and to create a perspective for seamless integration of the stepsize control with basic adaptive filtering, conveniently, the first-order linear Markov model \cite{Haykin2002} was then proposed to quantify our time-varying system \cite{Enzner2006,Enzner2010},
\begin{equation}
\label{states}  
\VEC{h}(n\!+\!1)  \,=\,  a\cdot\VEC{h}(n) + \VEC{\Delta h}(n)\;, 
\end{equation}
with $\VEC{h}(n)$ referred to as the acoustic system state. Two consecutive states at time indices $n$ and $n\!+\!1$ are coupled by the \textit{time-invariant} transition coefficient $0\leq a \leq 1$, but the independent
process noise quantity $\VEC{\Delta h}(n)$ with generally time-varying $N\times N$ covariance matrix $\MATG{\Sigma_\Delta}(n)=\EV{\VEC{\Delta h}(n)\VEC{\Delta h}^T(n)}$ yet changes the current system state into an unpredictable direction. This formal model clearly constitutes a workable simplification over the real world. Still it resembles the common observation that the echo path impulse response is almost time-invariant as long as the acoustic enclosure is fixed, i.e., small vector norm $||\VEC{\Delta h}(n)||$, but it continuously changes its shape in the case of natural interaction of the user with the acoustic enclosure, i.e., larger $||\VEC{\Delta h}(n)||$, as shown by the experimentation in \cite{Enzner2006}.

Eqs.~(\ref{measurements}) and (\ref{states}) then together
form a linear stochastic state-space model of the unknown system state $\VEC{h}(n)$. Further assuming independent and temporally uncorrelated process and observation noises, an estimate of the latent system state at time $n$, in this case, is known to be computed elegantly by the discrete-time Kalman filter (KF). It consists of the following set of recursively and iteratively coupled matrix equations \cite{Scharf91,Haykin2002}:
\color{black}
\begin{eqnarray}
\label{kalman1}
\widehat{\VEC{h}}(n) & \!\!\!=\!\!\! & a\cdot \widehat{\VEC{h}}^+(n)   \\[0.25ex]
\label{kalman3}  
\widehat{\VEC{h}}^+(n) &  \!\!\!=\!\!\! & \widehat{\VEC{h}}(n-1) + \VEC{k}(n)\, e(n)\quad  
        \\
        \label{errorKF}
e(n)  &  \!\!\!=\!\!\! & y(n) - \VEC{x}^T(n) \widehat{\VEC{h}}(n-1) 
\\[0.25ex]
\label{kalman5}  
\VEC{k}(n) & \!\!\!=\!\!\! & \MAT{p}(n)\VEC{x}(n)
\Bigl( \VEC{x}^T(n)\MAT{p}(n)\VEC{x}(n) + \sigma_{s+n}^2(n) \Bigr)^{-1} \\
\label{kalman2}
\VEC{p}(n\!+\!1) & \!\!\!=\!\!\! & a^2\cdot \VEC{p}^+(n) +  
\MATG{\Sigma_\Delta}(n)\\[0.25ex]
\label{kalman4}		
\VEC{p}^+(n) &  \!\!\!=\!\!\! & \Bigl( \MAT{I} - \VEC{k}(n)\VEC{x}^T(n) \Bigr) \MAT{p}(n)
\end{eqnarray}
where (\ref{kalman1}) and (\ref{kalman3}) operate in a prediction-correction fashion to determine the state estimate $\widehat{\VEC{h}}(n)$. In doing so, the $N\times 1$ Kalman gain $\VEC{k}(n)$ in 
(\ref{kalman5}) is employed as a weight vector that depends on the $N\times N$ state error covariance matrix
$\VEC{p}(n)=\EV{(\VEC{h}(n)-\widehat{\VEC{h}}(n))({\bf h}(n)-\widehat{\VEC{h}}(n))^T}$, which is in turn found recursively by (\ref{kalman2}) and (\ref{kalman4}). The term $\sigma_{s+n}^2(n)$ eventually refers to a-priori information of a time-varying observation noise variance related to \eqref{measurements}.

There are various approaches to the KF solution. On the one hand, it can be derived as a conditional mean estimation of the latent system state with assumptions of \textit{Gaussianity} of process, observations, and initial process state \cite{Scharf91}. Otherwise, it can still be described as an MMSE estimator with preconditioning to a linear filtering operation in place of the Kalman gain \cite{Haykin2002}. The KF therefore represents a good approach to the sought system state from various perspectives, subject to our simple model of Markovian state evolution \eqref{states} in either case.
Compared to LMS, the Kalman gain with accurate information of the time-varying system distance $\VEC{p}(n)$ and observation noise variance $\sigma_{s+n}^2(n)$ can be considered as an optimal adaptive stepsize formula in the recursive learning procedure of the system state $\widehat{\bf h}(n)$. It is thus supposed to achieve optimally fast and yet robust adaptation. In this way, the KF can be considered as an ever sought unification of adaptive filtering and adaptation control in acoustic echo cancellation \cite{Benesty2001,Haensler2004}, but a problem of providing the accurate a-priori information for the KF remains \cite{yang2017frequency}.

The KF was long avoided in acoustic system identification. This can be attributed to high computational load of the fully populated matrix algebra and to a risk of numerical instability in the case of higher-order adaptive filters \cite{Haykin2002}. Furthermore, a comprehensive signal model for the Kalman filter, particularly the aforementioned availability of accurate observation and process noise covariances of the acoustic state-space model in (\ref{measurements}) and (\ref{states}) seemed to be out of sight \cite{Haensler2004}. A general Kalman filtering approach employing observation overlap and practical considerations regarding the model covariances was reported in \cite{Paleologu2013}. However, it should be noted that the aforementioned traditional KF assumption of a temporally uncorrelated observation noise process is violated by overlapping observations. 

\subsection{Frequency-Domain Adaptive Kalman Filter (FDKF)}
\label{sec:FDKF}

It was the FDKF framework which overcame former limitations of KF theory for acoustic echo path tracking. To achieve decorrelation of the near-end signals, i.e., $s(n)$ and $n(n)$, and to reduce  the computational complexity of a KF implementation, an acoustic state-space model was described in the frame-wise discrete Fourier transform (DFT) domain \cite{Enzner2006}. In this way, we can create almost uncorrelated \textit{frames} of the near-end speech and observation noise, $s(n)+n(n)$, in the typical range of frame-based speech enhancement \cite{Vary2006} and thus meet the requirement of uncorrelatedness in the KF. Moreover, it was demonstrated that model covariances of the full-fledged KF in the DFT domain can be approximated very accurately and efficiently by the covariance diagonalization property in the DFT domain. This results in a computationally efficient narrowband treatment (i.e., frequency bins processed largely independently from each other by FDKF) in contrast to the fullband KF in (\ref{kalman1}) to (\ref{kalman4}). The original verbalism of \textit{block frequency-domain adaptive Kalman filtering} (FDKF) was converted into \textit{state-space frequency-domain adaptive filtering} (SSFDAF) \cite{Malik2011}
to highlight its strong relationship with seminal \textit{frequency-domain adaptive filtering} \cite{Ferrara80}.

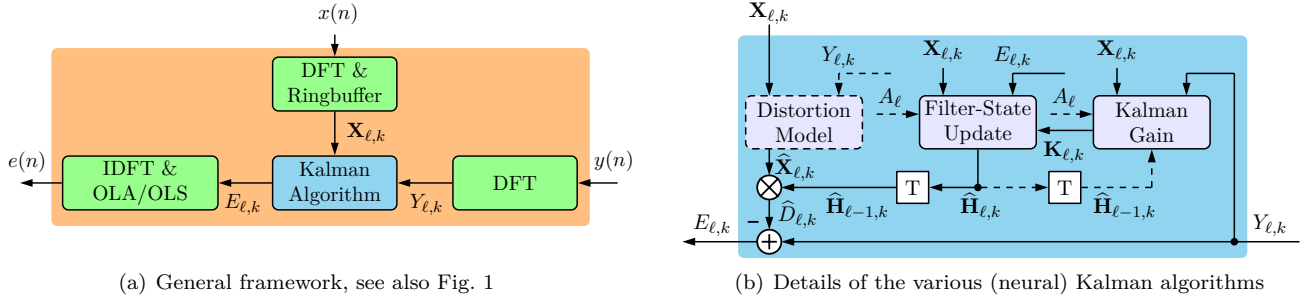
\begin{figure}[t]
    \subfigure[General framework, see also Fig. \ref{fig:systemlvl}]{
    \begin{minipage}{0.49\columnwidth}
        	\usetikzlibrary{dsp}
	\usetikzlibrary{chains}
	\usetikzlibrary{shapes.geometric}
	\usetikzlibrary{calc}

\begin{tikzpicture}[scale = 1.2]

\tikzstyle{encdec}=[trapezium, trapezium angle=67.5, draw, inner ysep=5pt, outer sep=0pt,text width=1, minimum height=10, line width=.8pt, fill=white]

\pgfarrowsdeclare{dsparrow}{dsparrow}
{
	\arrowsize=0.40pt
	\advance\arrowsize by .5\pgflinewidth
	\pgfarrowsleftextend{-4\arrowsize}
	\pgfarrowsrightextend{4\arrowsize}
}
{
	\arrowsize=0.40pt
	\advance\arrowsize by .5\pgflinewidth
	\pgfsetdash{}{0pt} 
	\pgfsetmiterjoin	 
	\pgfsetbuttcap		 
	\pgfpathmoveto{\pgfpoint{-4\arrowsize}{2.5\arrowsize}}
	\pgfpathlineto{\pgfpoint{4\arrowsize}{0pt}}
	\pgfpathlineto{\pgfpoint{-4\arrowsize}{-2.5\arrowsize}}
	\pgfpathclose
	\pgfusepathqfill
}

\pgfmathsetmacro{\toprow}{3.0}
\pgfmathsetmacro{\bottomrow}{0.9}
\pgfmathsetmacro{\middlerow}{(2.30}
\pgfmathsetmacro{\boxheight}{2.6}
\pgfmathsetmacro{\boxwidth}{6.0}

    \draw (3.75, \middlerow-0.75) node    (backdrop)     [rectangle, rounded corners=0.1cm, minimum width=26 em, minimum height=8.6 em, fill=orange!50] {}
            
            (3.95, \toprow)    coordinate  (StopX) {}
            
            (-0.5, \bottomrow) coordinate  (OutE)  {}
            (3.95, \bottomrow) node        (GS) [dspfilter, rounded corners=0.1cm, minimum width=\boxwidth em, minimum height=\boxheight em, fill=lightblue]{}
            (3.95, \bottomrow) node (t1) [align=center, text width=\boxwidth em] {Kalman Algorithm}

            (3.95, \middlerow) node (DFTx) [dspfilter, rounded corners=0.1cm, minimum width=\boxwidth em, minimum height=\boxheight em, align=center, fill=lightgreen] {} 
            (3.95, \middlerow) node (t1) [align=center, text width=\boxwidth em] {DFT \& \\ Ringbuffer}

            (6.5, \bottomrow) node (DFTy) [dspfilter, rounded corners=0.1cm, minimum width=\boxwidth em, minimum height=\boxheight em, align=center, fill=lightgreen] {} 
            (6.5, \bottomrow) node (t1) [align=center, text width=\boxwidth em] {DFT}
            
            (1.2, \bottomrow) node (Synth) [dspfilter, rounded corners=0.1cm, minimum width=7.5 em, minimum height=\boxheight em, align=center, fill=lightgreen] {} 
            (1.2, \bottomrow) node (t1) [align=center, text width=7 em] {IDFT \& \\ OLA/OLS}
            
            (7.95, \bottomrow) coordinate  (Mic)  {}
            (8.11, \bottomrow) coordinate  (InM)  {};

    \draw (4, \toprow+0.3) node (xt) {$x(n)$};
    \draw (-0.4, \bottomrow+0.3) node (et) {$e(n)$};
    \draw (7.9, \bottomrow+0.32) node (yt) {$y(n)$};

    \draw (4.4, \middlerow-0.7) node (Xt) {$\mathbf{X}_{\ell,k}$};
    \draw (2.65, \bottomrow-0.3) node (Et) {${E}_{\ell,k}$};
    \draw (5.25, \bottomrow-0.3) node (Yt) {${Y}_{\ell,k}$};

    \begin{scope}[start chain]

        \chainin (Synth);
        \chainin (OutE) [join=by dspconn];

        \chainin (Mic);
        \chainin (DFTy) 	[join=by dspconn];
        \chainin (GS) 	[join=by dspconn];
        \chainin (Synth) 	[join=by dspconn];

        \chainin (StopX);
        \chainin (DFTx) 	[join=by dspconn];
        \chainin (GS) 	[join=by dspconn];

    \end{scope}
	
\end{tikzpicture}
    	\label{fig:toplvl}
        \vspace{2mm}
    \end{minipage}
    }
    \hfill
    \subfigure[Details of the various (neural) Kalman algorithms]{
    \begin{minipage}{0.49\columnwidth}
        \hspace{-3mm}
        	\usetikzlibrary{dsp}
	\usetikzlibrary{chains}
	\usetikzlibrary{shapes.geometric}
	\usetikzlibrary{calc}

\begin{tikzpicture}[scale = 0.9]

\pgfarrowsdeclare{dsparrow}{dsparrow}
{
	\arrowsize=0.40pt
	\advance\arrowsize by .5\pgflinewidth
	\pgfarrowsleftextend{-4\arrowsize}
	\pgfarrowsrightextend{4\arrowsize}
}
{
	\arrowsize=0.40pt
	\advance\arrowsize by .5\pgflinewidth
	\pgfsetdash{}{0pt} 
	\pgfsetmiterjoin	 
	\pgfsetbuttcap		 
	\pgfpathmoveto{\pgfpoint{-4\arrowsize}{2.5\arrowsize}}
	\pgfpathlineto{\pgfpoint{4\arrowsize}{0pt}}
	\pgfpathlineto{\pgfpoint{-4\arrowsize}{-2.5\arrowsize}}
	\pgfpathclose
	\pgfusepathqfill
}

\tikzstyle{encdec}=[trapezium, trapezium angle=67.5, draw, inner ysep=5pt, outer sep=0pt,text width=1, minimum height=10, line width=.8pt, fill=white]

\pgfmathsetmacro{\toprow}{3.0}
\pgfmathsetmacro{\bottomrow}{1}
\pgfmathsetmacro{\middlerow}{(2.10}
\pgfmathsetmacro{\boxheight}{2}
\pgfmathsetmacro{\boxwidth}{5.5}
\pgfmathsetmacro{\hscale}{1.6}
\pgfmathsetmacro{\vscale}{1}

    \draw   (-1.83*\hscale, -1.75*\vscale) node    (backdrop)     [rectangle, rounded corners=0.1cm, minimum width=24 em, minimum height=10.5 em, fill=lightblue] {}

            (1.7*\hscale, -3.5*\vscale) coordinate (y) {}

            (0.95*\hscale, -0.4*\vscale) coordinate (c3) {} 
            (0.4*\hscale, -0.4*\vscale) coordinate (c4) {} 
            (-0.4*\hscale, -0.3*\vscale) coordinate (c5) {} 

            (-1.0*\hscale, -0.4*\vscale) coordinate (c3_2) {} 
            (-1.6*\hscale, -0.4*\vscale) coordinate (c4_2) {} 
            (-2.4*\hscale, -0.3*\vscale) coordinate (c5_2) {} 
            (-3.0*\hscale, -0.4*\vscale) coordinate (c3_3) {} 
            (-3.6*\hscale, -0.4*\vscale) coordinate (c4_3) {}
            (-4.4*\hscale, 0.5*\vscale) coordinate (c5_3) {} 

            (-4*\hscale, -1.3*\vscale) node (ref) [dspfilter, dashed, minimum width=\boxwidth em, rounded corners=0.1cm, minimum height=2.5 em, fill=blue!10] {}
            (-4*\hscale, -1.3*\vscale) node (t1) [align=center, text width=\boxwidth em] {Distortion \\ Model} 
            (-4*\hscale, -2.8*\vscale) coordinate (c6) {}

            (-2*\hscale, -1.3*\vscale) node (filt) [dspfilter, minimum width=\boxwidth em, rounded corners=0.1cm, minimum height=2.5 em, fill=blue!10] {}
            (-2*\hscale, -1.3*\vscale) node (t1) [align=center, text width=\boxwidth em] {Filter-State \\ Update} 
            
            (0.0*\hscale, -1.3*\vscale) node (gain) [dspfilter, minimum width=\boxwidth em, rounded corners=0.1cm, minimum height=2.5 em, fill=blue!10] {}
            (0.0*\hscale, -1.3*\vscale) node (t1) [align=center, text width=\boxwidth em] {Kalman \\ Gain} 
            
            (-4.4*\hscale, -2.5*\vscale) node  (M1In) [circle, inner sep= 4, fill=white]  {}
            (-4.4*\hscale, -2.5*\vscale) node (M1) [dspmixer] {} 
            (-4.4*\hscale, -3.5*\vscale) node  (AIn) [circle, inner sep= 4, fill=white]  {}
            (-4.4*\hscale, -3.5*\vscale) node (A1) [dspadder] {}

            (+0.1*\hscale, -0.4*\vscale) coordinate (c1) {} 
            (-4.4*\hscale, -0.4*\vscale) coordinate (c2) {} 

            (-2*\hscale, -2.5*\vscale) node (p3) [dspnodefull] {} 
            (+0.0*\hscale, -2.5*\vscale) coordinate (c7) {} 

            (0.95*\hscale, -3.5*\vscale) node (p2) [dspnodefull] {} 

            (-5.4*\hscale, -3.5*\vscale) coordinate (e) {}

            (-1*\hscale, -2.5*\vscale) node (T) [dspfilter, minimum width=1.5 em, minimum height=1.5 em, fill=white] {T}
            (-2.75*\hscale, -2.5*\vscale) node (T2) [dspfilter, minimum width=1.5 em, minimum height=1.5 em, fill=white] {T}
            
            ;
    
    \draw ([xshift=0mm, yshift=+3.0mm] c5.east) node (X_t) {${\mathbf{X}}_{\ell,k}$};
    \draw ([xshift=0mm, yshift=+3.0mm] c5_2.east) node (X_t) {${\mathbf{X}}_{\ell,k}$};
    \draw ([xshift=0mm, yshift=+2.0mm] c5_3.east) node (X_t) {$\mathbf{X}_{\ell,k}$};
    \draw ([xshift=+6mm, yshift=+3.0mm] p2.east) node (Y_t) {${Y}_{\ell,k}$};

    \draw ([xshift=0mm, yshift=+3.0mm] c4_2.east) node (Y_t) {${E}_{\ell,k}$};
    \draw ([xshift=0mm, yshift=+3.0mm] c4_3.east) node (Y_t) {${Y}_{\ell,k}$};

    \draw ([xshift=+5.0mm, yshift=-5.5mm] filt.east) node (Y_t) {$\mathbf{K}_{\ell,k}$};
    \draw ([xshift=+5.0mm, yshift=+4.5mm] filt.east) node (Y_t) {${A}_{\ell}$};
    \draw ([xshift=+5.0mm, yshift=+4.5mm] ref.east) node (Y_t) {${A}_{\ell}$};

    \draw ([xshift=+7mm, yshift=+3.75mm] M1.west) node (X_t) {$\widehat{\mathbf{X}}_{\ell,k}$};
    
    \draw ([xshift=0mm, yshift=-3.5mm] p3.east) node (H_t) {$\widehat{\mathbf{H}}_{\ell,k}$};
    \draw ([xshift=-5mm, yshift=-3.5mm] c7.east) node (H1_t) {$\widehat{\mathbf{H}}_{\ell-1,k}$};
    \draw ([xshift=14mm, yshift=-3.5mm] M1.east) node (H1_t) {$\widehat{\mathbf{H}}_{\ell-1,k}$};
    \draw ([xshift=+3.0mm, yshift=+5.0mm] A1.east) node (D_t) {$\widehat{D}_{\ell,k}$};

    \draw ([xshift=-5mm, yshift=+3.0mm] A1.east) node (min) {\huge -};
    \draw ([xshift=+5.5mm, yshift=+3mm] e.north) node (E_t) {${E}_{\ell,k}$};

    \draw ([yshift=+7.1mm] M1) coordinate (xhat_low) {};

    \draw ([yshift=-5.1mm] c5) coordinate (x_low1) {};
    \draw ([yshift=-4.1mm] c4) coordinate (y_low1) {}; 
    \draw ([yshift=-5.1mm] c5_2) coordinate (x_low2) {};
    \draw ([yshift=-4.1mm] c4_2) coordinate (y_low2) {}; 
    \draw ([yshift=-13.1mm] c5_3) coordinate (x_low3) {};
    \draw ([yshift=-4.1mm] c4_3) coordinate (y_low3) {}; 

    \draw ([xshift=2.5mm, yshift=1.5mm] ref.east) coordinate (a1) {};
    \draw ([yshift=1.5mm] filt.west) coordinate (a2) {};
    \draw ([xshift=2.5mm, yshift=1.5mm] filt.east) coordinate (a3) {};
    \draw ([yshift=1.5mm] gain.west) coordinate (a4) {};

    \draw ([yshift=-1.5mm] filt.east) coordinate (k2) {};
    \draw ([yshift=-1.5mm] gain.west) coordinate (k1) {};

    \begin{scope}[start chain]

        \chainin (k1);
        \chainin (k2)       [join=by dspconn];
        
        \chainin (filt);
        \chainin (p3)       [join=by dspline];
        \chainin (T2)       [join=by dspconn];
        \chainin (M1)       [join=by dspconn];

        \chainin (xhat_low);
        \chainin (M1) 	[join=by dspconn];
        \chainin (A1) 	[join=by dspconn];

        \chainin (y);
        \chainin (A1) 	[join=by dspconn];
        \chainin (e) 	[join=by dspconn];

        \chainin (p2);
        \chainin (c3) 	[join=by dspline];
        \chainin (c4) 	[join=by dspline];
        \chainin (y_low1) 	[join=by dspconn];

        \chainin (c3_2) ;
        \chainin (c4_2) 	[join=by dspline];
        \chainin (y_low2) 	[join=by dspconn];
        
        \chainin (c5);
        \chainin (x_low1) 	[join=by dspconn];
        \chainin (c5_2);
        \chainin (x_low2) 	[join=by dspconn];
        \chainin (c5_3);
        \chainin (x_low3) 	[join=by dspconn];

        \begin{scope}[dashed]
            \chainin (c3_3) ;
            \chainin (c4_3) 	[join=by dspline];
            \chainin (y_low3) 	[join=by dspconn];

            \chainin (p3);
            \chainin (T) 	    [join=by dspconn];
            \chainin (c7) 	    [join=by dspline];
            \chainin (gain) 	[join=by dspconn];

            \chainin (a1) ;
            \chainin (a2) 	[join=by dspconn];
            \chainin (a3) ;
            \chainin (a4) 	[join=by dspconn];
        
        \end{scope}
        
    \end{scope}
	
\end{tikzpicture}
    	\label{fig:midlvl}
        \vspace{2mm}
    \end{minipage}
    }
    \caption{Overview of a (neural) Kalman filter approach for acoustic echo cancellation. It operates in the discrete Fourier transform (DFT) domain and is constrained to overlap-add (OLA) or overlap-save (OLS) processing of the frequency bins. Notations are for the general case of multiple filter taps (bold fonts). Note that block ''T'' represents a delay unit.}
    \label{fig:overview}
    \vspace{-3.5mm}
\end{figure}

Figure~\ref{fig:overview} gives an overview of our AEC system definition based on FDKF and its extensions by means of neural network capabilities for nonlinear processing or covariance learning, with upper-case denoting frequency-domain signals. In the scope of this article, according to Fig.~\ref{fig:toplvl}, the sequences $x(n)$ and $y(n)$ are segmented into frames of length $K$ with frame shift $R$, followed by frame-wise windowing and $K$-point DFT. Using the frequency-domain representations $X_{\ell,k}$ and $Y_{\ell,k}$ of the respective signals at frame index $\ell$ and discrete frequency bin $k$, the Kalman algorithm is employed for obtaining an echo estimate $\widehat{D}_{\ell,k}$ and an error signal $E_{\ell,k}$ in the DFT domain by subtraction of $\widehat{D}_{\ell,k}$ from the microphone signal $Y_{\ell,k}$ .
The reconstruction of a time-domain sequence $e(n)$ relies on frame-wise IDFT and subsequent windowing for overlap-add (OLA) or overlap-save (OLS) synthesis. All methods implemented with OLA use a square-root Hann window. Note that this setup is not designed to suppress any background noise present at the microphone.

As shown by the detail of Fig.~\ref{fig:midlvl}, for echo estimation, the FDKF algorithm uses its estimated linear system state that---in the AEC task---acts as a single- or multi-tap adaptive filter \mbox{$\widehat{\VEC{H}}_{\ell,k}=\big(\widehat{H}_{\ell,k}^{(1)}, \widehat{H}_{\ell,k}^{(2)}, ..., \widehat{H}_{\ell,k}^{(L)}\big)^T$} per frequency bin. It is applied to $L$ most recent frames of a reference signal  \mbox{$\widehat{\VEC{X}}_{\ell,k}=\big(\widehat{X}_{\ell,k}, \widehat{X}_{\ell-1,k}, ..., \widehat{X}_{\ell-L+1,k}\big)^T$} to form an echo estimate $\widehat{D}_{\ell,k} = \widehat{\VEC{X}}_{\ell,k}^T \cdot \widehat{{\VEC{H}}}_{\ell-1,k}$ and a related error signal $E_{\ell,k}=Y_{\ell,k}-\widehat{D}_{\ell,k}$ by subtraction in the DFT domain.
The general multi-tap filter case formally includes the case of a single frame tap where needed for brevity.

Moreover, Fig.~\ref{fig:midlvl} shows the entire FDKF with its DNN-based extensions from pp.\ \pageref{sec:Hybrid}ff.\ broken down into three core blocks. The ''Filter-State Update'' block computes a misalignment of its previous filter state $\widehat{\VEC{H}}_{\ell-1,k}$ w.r.t.\ the microphone signal and returns, by analogy with (\ref{kalman1}) to (\ref{errorKF}), an updated filter state in the DFT domain \cite{Enzner2006,Jung2014,Kuech2014}
\begin{equation}
    \widehat{\VEC{H}}_{\ell,k} = A\, \widehat{\VEC{H}}_{\ell-1,k} + 
    A\,
    {\VEC{K}}_{\ell,k} E_{\ell,k}\quad \text{with}\quad E_{\ell,k} = Y_{\ell,k} - {\widehat{\VEC{X}}_{\ell,k}^T \cdot \widehat{{\VEC{H}}}_{\ell-1,k}} \;,
    \label{eq:H_FDKF}
\end{equation}
where typically $0\ll A\leq1$ (close to unity) now is in the role of the state-transition factor in the DFT domain. Some extended algorithms may consider a frame-wise factor ${A}_{\ell}$ as shown in Fig.~\ref{fig:midlvl}. The analogy with (\ref{kalman1}) to (\ref{errorKF}) specifically lies in the identification of our frame time index $\ell$ here with the sampling time index $n$ there, while the FDKF idea consists in the largely independent processing of our frequencies $k$ here. The required weighting
\begin{align}
    {\VEC{K}}_{\ell,k} &= \frac{R}{K} {P}_{\ell,k} \VEC{X}^*_{\ell,k}  \cdot \left( \frac{R}{K} \left[ \VEC{X}^{T}_{\ell,k} {P}_{\ell,k} \VEC{X}_{\ell,k} \right] + {\Psi}^\mathrm{S+N}_{\ell,k} \right)^{-1}
    \label{eq:K_FDKF}
\end{align}
to control the amplitude of the update per time-frequency bin are delivered by the ''Kalman gain'' block in analogy with (\ref{kalman5}), whereby the required observation noise variance is empirically (i.e., beyond KF theory) computed as
\begin{equation}
        {\Psi}^\mathrm{S+N}_{\ell,k} = \beta {\Psi}^\mathrm{S+N}_{\ell-1}(k)
        + (1-\beta) |{E}_{\ell,k}|^2
\end{equation}
with smoothing factor $\beta=0.5$. A common state error covariance $P_{\ell,k}$ recurs in analogy to  (\ref{kalman2}) and (\ref{kalman4}) as
\begin{equation}
        {P}_{\ell+1,k} =  A^2\, {P}_{\ell,k} \Big(1-\frac{R}{K} \VEC{X}^{T}_{\ell,k} \VEC{K}_{\ell,k}\Big) +  {\Psi}_{\ell,k}^\triangle
\end{equation}
with the process noise covariance ${\Psi}_{\ell,k}^\triangle$ depending on the application.
A careful choice of a suitable constant is a good recommendation in the interest of an overall robust system, while more sophisticated approaches with potentially advanced performance are available \cite{Enzner2006,Paleologu2013,yang2017frequency}. For later experiments, we will specifically employ \cite{Jung2014, Seidel2024}
\begin{equation}
    \begin{split}
        \Psi_{\ell,k}^\triangle = \left({P}_{\ell,k} + |\widehat{H}_{\ell-1,k}|^2 \right) \cdot (1-A^2).
    \end{split}
    \label{eq:noise_cov}
\end{equation}

Eventually, the ''Distortion Model'' block, which is entirely beyond the linear KF framework, refers to an additional conversion of the original far-end signal $\VEC{X}_{\ell,k}$ into the actual input signal $\widehat{\VEC{X}}_{\ell,k}$ of the KF algorithm. The conversion may include simple preprocessing such as delay compensation or sophisticated nonlinear loudspeaker modeling. It potentially uses the microphone signal $Y_{\ell,k}$ to capture such nonlinearities.

\section{Hybrid Approaches to a Kalman AEC} 
\label{sec:Hybrid}

The FDKF is known for excellent preservation of the near-end speaker even in the double-talk (DT) condition \cite{Franzen2018a}, but its echo cancellation capability is limited by a number of factors. Major limitations lie with the model linearity and the difficulty of obtaining a time-varying a-priori acoustic state-space representation (i.e., the model covariances) for practical Kalman gain computation \cite{yang2017frequency}. This hampers the model's representation of nonlinear loudspeaker distortions and its rapid reconvergence towards steady-state performance.

As mentioned before, end-to-end neural network-based AEC methods have been widely proposed in research throughout the past years. While those approaches often achieve impressive echo suppression results, most of the time we see either big model footprints and/or a trade-off in near-end speech quality during double-talk.

\begin{table}[t]
    \vspace{0mm}
	\setlength{\tabcolsep}{.35em}
	\caption{Overview of original {\tt FDKF} \cite{Enzner2006,Franzen2018a} and related DNN-based approaches \cite{Haubner2024, Yang2023, Zhang2023,Zhang2022d}
 in terms of the three core functions from Fig.\,\ref{fig:midlvl} along with further aspects. Model-based components are labelled {\color{blue}SP}, while DNN-based ones are labelled {\color{red}DNN}. Note that we have $\widehat{{X}}^\mathrm{\textcolor{red}{DNN}}_{\ell,k} = \widehat{{X}}^\mathrm{\textcolor{red}{DNN}}_{\ell,k}({\mathbf{X}}_{\ell,k},{\mathbf{Y}}_{\ell,k})$ and $A_{\ell}^\mathrm{\textcolor{red}{DNN}} = A_{\ell}^\mathrm{\textcolor{red}{DNN}}({\mathbf{X}}_{\ell,k=0:K\!-\!1},{\mathbf{Y}}_{\ell,k=0:K\!-\!1})$.}
	\vspace{-1.0mm}
    \centering
	\newcolumntype{R}{>{\raggedleft\arraybackslash}X}
\newcolumntype{C}{>{\center\arraybackslash}X}
\newcommand{\bftab}{\fontseries{b}\selectfont}

\setlength{\tabcolsep}{7pt}
\begin{tabular}{l l l l l l l }

 \multirow{2}{*}{\textbf{Model}} & \multicolumn{3}{c}{\textbf{Blocks in Figure\,\ref{fig:midlvl}}} & \multirow{2}{*}{\vtop{\hbox{\strut{\textbf{\!\!State\ Tr.}}} \hbox{\strut{\textbf{\!\!Factor}}}}} & \multirow{2}{*}{\vtop{\hbox{\strut{\textbf{\!\!\!\!Filter}}} \hbox{\strut{\textbf{\!\!\!\!Taps}}}}} & \multirow{2}{*}{\vtop{\hbox{\strut{\textbf{OLA/}}} \hbox{\strut{\textbf{OLS}}}}}\\

 \cmidrule(lr){2-4}

 & \textbf{Dist. Model}\!\! & \textbf{\!\!Filter-State Update}\!\!\!\! & \textbf{\!\!Kalman Gain} & & \\

 \toprule

 {\tt FDKF} \cite{Enzner2006,Franzen2018a} 
 & ${{X}}_{\ell,k}$ 
 & \!\!$\widehat{{H}}^\mathrm{\textcolor{blue}{SP}}_{\ell,k}({{E}}_{\ell,k},{{X}}_{\ell,k}, {{K}}^\mathrm{SP}_{\ell,k})$ 
 & \!\!${{K}}^\mathrm{\textcolor{blue}{SP}}_{\ell,k}({{X}}_{\ell,k},{{Y}}_{\ell,k}, \widehat{{H}}^\mathrm{SP}_{\ell-1,k})$  
 & \!$A=0.998$ 
 & \!\!\!\!single & OLS\\
 {\tt DLAC-Kalman} \cite{Haubner2024}\!\!\!\!\!\! 
 & ${{X}}_{\ell,k}$ 
 & \!\!$\widehat{\mathbf{H}}^\mathrm{\textcolor{blue}{SP}}_{\ell,k}({{E}}_{\ell,k},{{X}}_{\ell,k}, {\mathbf{K}}^\mathrm{DNN}_{\ell,k})$\!\!  
 & \!\!${\mathbf{K}}^\mathrm{\textcolor{red}{DNN}}_{\ell,k}({\mathbf{X}}_{\ell,k},{{Y}}_{\ell,k}, \widehat{\mathbf{H}}^\mathrm{SP}_{\ell-1,k})$\!\!\!\!\!\!\!\!  
 & \!$A=1$ 
 & \!\!\!\!multi & OLA\\
 {\tt NKF} \cite{Yang2023} 
 & ${{X}}_{\ell,k}$ 
 & \!\!$\widehat{\mathbf{H}}^\mathrm{\textcolor{blue}{SP}}_{\ell,k}({{E}}_{\ell,k},{{X}}_{\ell,k}, {\mathbf{K}}^\mathrm{DNN}_{\ell,k})$\!\!   
 & \!\!${\mathbf{K}}^\mathrm{\textcolor{red}{DNN}}_{\ell,k}({\mathbf{X}}_{\ell,k},{{Y}}_{\ell,k}, \widehat{\mathbf{H}}^\mathrm{SP}_{\ell-1,k})$\!\!\!\!\!\!\!\!   
 & \!$A=1$ 
 & \!\!\!\!multi & OLA\\
 {\tt NeuralKalman} \cite{Zhang2023}\!\!\!\!\!\! 
 & $\widehat{{X}}^\mathrm{\textcolor{red}{DNN}}_{\ell,k}$
 & \!\!$\widehat{{H}}^\mathrm{\textcolor{red}{DNN}}_{\ell,k}({{E}}_{\ell,k},{{X}}_{\ell,k}, {{K}}^\mathrm{SP}_{\ell,k})$\!\!   
 & \!\!${{K}}^\mathrm{\textcolor{blue}{SP}}_{\ell,k}({{X}}_{\ell,k},{{Y}}_{\ell,k}, \widehat{{H}}^\mathrm{DNN}_{\ell-1,k})$   
 & \!$A_{\ell}^\mathrm{\textcolor{red}{DNN}}\!$
 & \!\!\!\!single & OLS\\
  {\tt DeepAdaptive} \cite{Zhang2022d}\!\!\!\!\!\! 
 & $\widehat{{X}}^\mathrm{\textcolor{red}{DNN}}_{\ell,k}$
 & \!\!$\widehat{\mathbf{H}}^\mathrm{\textcolor{blue}{SP}}_{\ell,k}({{E}}_{\ell,k},{\widehat{X}}_{\ell,k}, {\mathbf{K}}^\mathrm{DNN}_{\ell,k})$\!\!   
 & \!\!${\mathbf{K}}^\mathrm{\textcolor{red}{DNN}}_{\ell,k}(\widehat{\mathbf{X}}_{\ell,k},{{Y}}_{\ell,k})$   
 & \!$A=1$ 
 & \!\!\!\!multi & OLA\\

 \end{tabular}
	\label{tab:diff}
\end{table}

The more recent works on hybrid AEC models attempt to combine the strengths of both classical and DNN-based approaches. Specifically, the {\tt FDKF} framework leverages {its} good echo cancellation without NE degradation, while not too complex neural networks augment {\tt FDKF} to support faster convergence, better handling on nonlinearities, and thus overall larger steady-state echo reduction. We refer to these approaches as \textit{neural Kalman filters}.

Some representative methods of the emerging trend of neural Kalman filtering are summarized in Table\,\ref{tab:diff} alongside their particular DNN-based treatment of the core blocks from Fig.\ \ref{fig:midlvl} (marked by the label ''{\color{red}DNN}'') in functional form. Wherever algorithms are marked with the label ''{\color{blue}SP}'', they follow the classical {\tt FDKF} algorithm. Most of the approaches rely directly on the far-end reference signal ${\mathbf{X}}_{\ell,k}$ for echo estimation, with the exception of {\tt NeuralKalman}~\cite{Zhang2023} and {\tt DeepAdaptive}~\cite{Zhang2022d}, which apply 
a learned distortion $\widehat{\mathbf{X}}^\mathrm{DNN}_{\ell,k}$, supposedly for nonlinear loudspeaker modeling. 
{\tt NeuralKalman} exclusively is the one to also replace the filter-state update block by a supporting DNN. The deep learning-based adaptation control inspired by Kalman theory ({\tt DLAC-Kalman})~\cite{Haubner2024}, the {\tt DeepAdaptive}~\cite{Zhang2022d}, and the neural Kalman filtering ({\tt NKF})~\cite{Yang2023} propose different architectures for DNN-based estimation of the Kalman gain to improve the (re)convergence of AEC, {while their state transition factor can simply be chosen as $A_{\ell}=A=1$.}
Noteworthy, the {\tt NeuralKalman}~\cite{Zhang2023} approach additionally uses a frame-wise DNN-based $A_{\ell}^\mathrm{DNN}$ for improved filter-state update upon RIR changes, but in turn relies on a model-based Kalman gain formula. The reported filter taps in Table\,\ref{tab:diff} follow the original articles, while single-tap systems could easily be implemented multi-tap and vice-versa.
Note that we replicate the authors' methods as closely as possible. Our {\tt NeuralKalman} implementation includes all submodules proposed in \cite{Zhang2023}. For {\tt DLAC-Kalman}, we choose the narrowband implementation (NB-DNN)~\cite{Haubner2024} with magnitude input features $|Y_{\ell,k}|$, $|X_{\ell,k}|$, $|E_{\ell-1,k}|$, and $|\widehat{D}_{\ell-1,k}|$, including their input feature normalization. The {\tt NKF} model is adapted from the authors' Github code as referenced in \cite{Yang2023}, while {\tt DeepAdaptive} follows the ''Proposed (no NR)'' model \cite{Zhang2022d}.

Another distinction between neural Kalman filter models not shown in Table\,\ref{tab:diff} lies in the processing of DNN input vectors. For {\tt NKF} and our {\tt DLAC} implementation (chosen due to its similarity with {\tt FDKF}), frequency bins are processed independently from each other by a single DNN, which represents weight sharing of frequency-individual DNNs. For {\tt NeuralKalman} and {\tt DeepAdaptive}, frequency bins are processed jointly due to the use of fully connected layers with frequency vector input.
This choice may not only have an effect on performance (as demonstrated in \cite{Haubner2024}), but also on computational complexity and memory footprint. Per-bin processing with shared weights usually allows for smaller networks with a lower---and more importantly, DFT size-independent---parameter count, but 
in turn comes with higher computational complexity as the number of nodes in a network typically increases after the input layer. For per-bin processing, the respective computations have to be performed $K/2+1$ times and are not shared as it is the case with joint frequency-bin processing.

\subsection{DNN-Based Kalman Gain Estimation}
\label{ssec:Gain_DNN}

A common research trend in hybrid modeling, be it NLMS- or FDKF-based, focuses mainly on the support of stepsize or Kalman gain estimation by a DNN. In the {\tt FDKF}, the Kalman gain controls the update of the acoustic system estimate as described in \eqref{eq:K_FDKF}, specifically regarding the convergence speed and accuracy. Typical FDKF implementations without DNN pursue a smooth update strategy to preserve the NE signal and avoid erratic system state changes, but would require a better rate of convergence and larger steady-state echo cancellation.

Three of the presented methods in Tab.\ \ref{tab:diff} employ a DNN for estimation of the Kalman gain, each with their own approach to the task. The most straight-forward method is the {\tt NKF}~\cite{Yang2023} which employs a DNN consisting of only a few fully connected and GRU layers to directly estimate ${\VEC{K}}_{\ell,k}$. Due to the fact that a very small network is applied to each time-frequency bin individually---using the same weights---this results in a very efficient solution with a small parameter footprint. The {\tt DeepAdaptive} model~\cite{Zhang2022d}, on the other hand, employs a DNN (mainly consisting of LSTM layers) which is fully connected between all input features, whereby most of the network (except the final, fully connected layer) is shared with the distortion model, see Fig.\ \ref{fig:NKal} on the right. Here, instead of estimating ${\VEC{K}}_{\ell,k}$ directly, the model outputs the (single-tap) stepsize {${{\mu}}^\mathrm{DNN}_{\ell,k}$} of a ''frequency-domain NLMS'' algorithm, corresponding to an effective (multi-tap) Kalman gain {${\VEC{K}}^\mathrm{DNN}_{\ell,k} = \frac{{{\mu}}^\mathrm{DNN}_{\ell,k}({{X}}_{\ell,k},{{Y}}_{\ell,k})}{\widehat{\VEC{X}}_{\ell,k}^T \widehat{\VEC{X}}_{\ell,k}} \widehat{\VEC{X}}_{\ell,k}^*$}. 
This approach therefore still follows a model-based update rule, but replaces components  ${{\mu}}_{\ell,k}$ and ${\VEC{X}}_{\ell,k}$ with DNN-based estimations and removes the necessity of estimating a-priori acoustic state-space representations.
The third approach, {\tt DLAC-Kalman}, goes even further by estimating two variables ${m}_{\ell,k}^\mathrm{DNN-\mu}$ and ${m}_{\ell,k}^\mathrm{DNN-e}$, which are then used to calculate the Kalman gain (based on the stepsize as defined in \cite{Haubner2024})
\begin{equation}
    {{\VEC{K}}^\mathrm{DNN}_{\ell,k}} = \frac{{m}_{\ell,k}^\mathrm{DNN-\mu}(|X_{\ell,k}|,|Y_{\ell,k}|,|E_{\ell-1,k}|,|\widehat{D}_{\ell-1,k}|)}
    {\psi_{\ell,k}+|{m}_{\ell,k}^\mathrm{DNN-e}(|X_{\ell,k}|,|Y_{\ell,k}|,|E_{\ell-1,k}|,|\widehat{D}_{\ell-1,k}|) \cdot{E}_{\ell,k}|^2 + \delta^{VSS}} {\VEC{X}}_{\ell,k}^*
\end{equation}
with average reference signal power $\psi_{\ell,k} = \lambda \psi_{\ell-1,k} + (1-\lambda) {\VEC{X}}_{\ell,k}^T {\VEC{X}}_{\ell,k}$, parameter $\lambda = 0.9$, and a small regularization constant $\delta^{VSS} > 0$. Please note that the symbol $E_{\ell,k}$ is missing in the original article \cite{Haubner2024}, although intended by the authors. Similar to {\tt NKF}, the {\tt DLAC-Kalman} DNN only consists of a few fully connected and GRU layers.

\subsection{DNN-Based State Transition Estimation}
\label{ssec:Transition_DNN}

The state transition factor $A$ controls the retention of the {\tt FDKF}'s previous state and thereby directly affects the model's reconvergence capabilities. This value is usually fixed close to unity, which allows the model to retain most of its system state during phases without far-end excitation. However, this means that on a significant change in the acoustic system, the {\tt FDKF} has to actively erase its previous estimate $\widehat{\VEC{H}}_{\ell,k}$ through filter updates. Adaptive control over the $A$ parameter may significantly support reconvergence by ''forgetting'' previous filter estimates.

\begin{figure}[t]
    \centering
    \begin{minipage}{0.56\columnwidth}
        	\usetikzlibrary{dsp}
	\usetikzlibrary{chains}
	\usetikzlibrary{shapes.geometric}
	\usetikzlibrary{calc}

\begin{tikzpicture}[scale = 0.9]

\pgfarrowsdeclare{dsparrow}{dsparrow}
{
	\arrowsize=0.40pt
	\advance\arrowsize by .5\pgflinewidth
	\pgfarrowsleftextend{-4\arrowsize}
	\pgfarrowsrightextend{4\arrowsize}
}
{
	\arrowsize=0.40pt
	\advance\arrowsize by .5\pgflinewidth
	\pgfsetdash{}{0pt} 
	\pgfsetmiterjoin	 
	\pgfsetbuttcap		 
	\pgfpathmoveto{\pgfpoint{-4\arrowsize}{2.5\arrowsize}}
	\pgfpathlineto{\pgfpoint{4\arrowsize}{0pt}}
	\pgfpathlineto{\pgfpoint{-4\arrowsize}{-2.5\arrowsize}}
	\pgfpathclose
	\pgfusepathqfill
}

\tikzstyle{encdec}=[trapezium, trapezium angle=67.5, draw, inner ysep=5pt, outer sep=0pt,text width=1, minimum height=10, line width=.8pt, fill=white]

\pgfmathsetmacro{\toprow}{3.0}
\pgfmathsetmacro{\bottomrow}{1}
\pgfmathsetmacro{\middlerow}{(2.10}
\pgfmathsetmacro{\boxheight}{2}
\pgfmathsetmacro{\boxwidth}{5.5}
\pgfmathsetmacro{\hscale}{1.6}
\pgfmathsetmacro{\vscale}{1}

    \draw   (-2.05*\hscale, -0.15*\vscale) node    (backdrop)     [rectangle, rounded corners=0.1cm, minimum width=25.5 em, minimum height=18.75 em, fill=lightblue] {}

            (1.7*\hscale, -3.5*\vscale) coordinate (y) {}

            (0.95*\hscale, -0.4*\vscale) coordinate (c3) {} 
            (0.4*\hscale, -0.4*\vscale) coordinate (c4) {} 
            (-0.0*\hscale, -0.3*\vscale) coordinate (c5) {} 

            (-1.0*\hscale, -0.4*\vscale) coordinate (c3_2) {} 
            (-2*\hscale, -0.4*\vscale) coordinate (c4_2) {} 
            (-2.4*\hscale, -0.4*\vscale) coordinate (c5_2) {} 
            (-3.0*\hscale, -0.4*\vscale) coordinate (c3_3) {} 
            (-1.1*\hscale, 3.7*\vscale) coordinate (c5_3) {}
            (-4.4*\hscale, 3.7*\vscale) coordinate (c4_3) {} 

            (-3*\hscale, 3.7*\vscale) coordinate (y_in_2) {} 

            (-3.2*\hscale, -0.4*\vscale) coordinate (x_in_1) {}

            (-2*\hscale, +2.75*\vscale) node (she) [dspfilter, minimum width=2.2*\boxwidth em, rounded corners=0.1cm, minimum height=2.5 em, fill=red!20] {}
            (-2*\hscale, +2.75*\vscale) node (t1) [align=center, text width=2.4*\boxwidth em] {Shared Feature \\ Encoder} 
            (-2*\hscale, +1.9*\vscale) node (she_m) [dspnodefull] {}
            (-1*\hscale, +1.9*\vscale) coordinate (she_r) {}
            (-3*\hscale, +1.9*\vscale) coordinate (she_l) {}
            (-1*\hscale, +1.5*\vscale) coordinate (she_r_) {}
            (-3*\hscale, +1.5*\vscale) coordinate (she_l_) {}

            (-1*\hscale, +1.0*\vscale) node (tm) [dspfilter, minimum width=\boxwidth em, rounded corners=0.1cm, minimum height=2.5 em, fill=red!20] {}
            (-1*\hscale, +1.0*\vscale) node (t1) [align=center, text width=\boxwidth em] {Transition \\ Model} 
            (-1*\hscale, +0.0*\vscale) node (tm_m) [dspnodefull] {}
            (-0.4*\hscale, +0.0*\vscale) coordinate (tm_r) {}
            (-1.6*\hscale, +0.0*\vscale) coordinate (tm_l) {}
            (-0.4*\hscale, -0.8*\vscale) coordinate (tm_r_) {}
            (-1.6*\hscale, -0.8*\vscale) coordinate (tm_l_) {}

            (-3.0*\hscale, -0.0*\vscale) coordinate (M0_pre)

            (-3*\hscale, +1.0*\vscale) node (ref) [dspfilter, minimum width=\boxwidth em, rounded corners=0.1cm, minimum height=2.5 em, fill=red!20] {}
            (-3*\hscale, +1.0*\vscale) node (t1) [align=center, text width=\boxwidth em] {Distortion \\ Model} 
            (-3*\hscale, -2.8*\vscale) coordinate (c6) {}

            (-2*\hscale, -1.3*\vscale) node (filt) [dspfilter, minimum width=\boxwidth em, rounded corners=0.1cm, minimum height=2.5 em, left color=red!20, right color=blue!10,shading angle=135] {}
            (-2*\hscale, -1.3*\vscale) node (t1) [align=center, text width=\boxwidth em] {Filter-State \\ Update} 
            
            (0.0*\hscale, -1.3*\vscale) node (gain) [dspfilter, minimum width=\boxwidth em, rounded corners=0.1cm, minimum height=2.5 em, fill=blue!10] {}
            (0.0*\hscale, -1.3*\vscale) node (t1) [align=center, text width=\boxwidth em] {Kalman \\ Gain} 
            
            (-4.4*\hscale, -2.5*\vscale) node  (M1In) [circle, inner sep= 4, fill=white]  {}
            (-4.4*\hscale, -2.5*\vscale) node (M1) [dspmixer] {} 
            (-4.4*\hscale, -3.5*\vscale) node  (AIn) [circle, inner sep= 4, fill=white]  {}
            (-4.4*\hscale, -3.5*\vscale) node (A1) [dspadder] {}

            (-4.4*\hscale, 0.0*\vscale) node (M0) [dspfilter, minimum width=\boxwidth em, rounded corners=0.1cm, minimum height=2.5 em, fill=blue!10] {}
            (-4.4*\hscale, 0.0*\vscale) node (M0_t) [align=center, text width=\boxwidth em] {Deep \\ Filtering}

            (+0.1*\hscale, -0.4*\vscale) coordinate (c1) {} 
            (-4.4*\hscale, -0.4*\vscale) coordinate (c2) {} 

            (-2*\hscale, -2.5*\vscale) node (p3) [dspnodefull] {} 
            (+0.0*\hscale, -2.5*\vscale) coordinate (c7) {} 

            (0.95*\hscale, -3.5*\vscale) node (p2) [dspnodefull] {} 

            (-5.7*\hscale, -3.5*\vscale) coordinate (e) {}

            (-1*\hscale, -2.5*\vscale) node (T) [dspfilter, minimum width=1.5 em, minimum height=1.5 em, fill=white] {T}
            (-3*\hscale, -2.5*\vscale) node (T2) [dspfilter, minimum width=1.5 em, minimum height=1.5 em, fill=white] {T}
            
            ;

    \draw ([xshift=5mm, yshift=+1.0mm] M0_pre.east) node (Y_t) {$\mathbf{M}_{\ell,k}$};
    
    \draw ([xshift=0mm, yshift=+3.0mm] c5.east) node (X_t) {${{X}}_{\ell,k}$};
    \draw ([xshift=0mm, yshift=+2.0mm] c5_3.east) node (X_t) {$\mathbf{X}_{\ell,k}$};
    \draw ([xshift=+6mm, yshift=+3.0mm] p2.east) node (Y_t) {${Y}_{\ell,k}$};

    \draw ([xshift=0mm, yshift=+2.0mm] y_in_2.east) node (X_t) {$\mathbf{Y}_{\ell,k}$};

    \draw ([xshift=3mm, yshift=-3.0mm] x_in_1.east) node (X_t) {${X}_{\ell,k}$};
    
    \draw ([xshift=0mm, yshift=+3.0mm] c4_2.east) node (Y_t) {${E}_{\ell,k}$};
    \draw ([xshift=0mm, yshift=+2.0mm] c4_3.east) node (Y_t) {$\mathbf{Y}_{\ell,k}$};

    \draw ([xshift=+5.0mm, yshift=-3.5mm] filt.east) node (Y_t) {${K}_{\ell,k}$};
    \draw ([xshift=+5.0mm, yshift=+9.5mm] filt.east) node (Y_t) {${A}_{\ell}$};

    \draw ([xshift=+7mm, yshift=+12.5mm] M1.west) node (X_t) {${\widehat{X}}_{\ell,k}$};

    \draw ([xshift=-27.5mm, yshift=+3.5mm] p3.east) node (H_t) {$\widehat{{H}}_{\ell-1,k}$};
    \draw ([xshift=-0mm, yshift=-3.5mm] p3.east) node (H_t) {$\widehat{{H}}_{\ell,k}$};
    \draw ([xshift=-5mm, yshift=-3.5mm] c7.east) node (H1_t) {$\widehat{{H}}_{\ell-1,k}$};
    \draw ([xshift=+3.0mm, yshift=+5.0mm] A1.east) node (D_t) {$\widehat{D}_{\ell,k}$};

    \draw ([xshift=-5mm, yshift=+3.0mm] A1.east) node (min) {\huge -};
    \draw ([xshift=+4.5mm, yshift=+3mm] e.north) node (E_t) {${E}_{\ell,k}$};

    \draw ([yshift=+7.1mm] M1) coordinate (xhat_low) {};

    \draw ([yshift=-5.1mm] c5) coordinate (x_low1) {};
    \draw ([yshift=-4.1mm] c4) coordinate (y_low1) {}; 
    \draw ([yshift=-4.1mm] c5_2) coordinate (x_low2) {};
    \draw ([yshift=-4.1mm] c4_2) coordinate (y_low2) {}; 
    \draw ([yshift=-4.51mm] c5_3) coordinate (x_low3) {};
    \draw ([yshift=-4.1mm] c4_3) coordinate (y_low3) {}; 

    \draw ([yshift=-4.5mm] y_in_2) coordinate (y_in_3) {}; 

    \draw ([xshift=2.5mm, yshift=1.5mm] ref.east) coordinate (a1) {};
    \draw ([yshift=1.5mm] filt.west) coordinate (a2) {};
    \draw ([xshift=2.5mm, yshift=1.5mm] filt.east) coordinate (a3) {};
    \draw ([yshift=1.5mm] gain.west) coordinate (a4) {};

    \draw ([yshift=-1.5mm] filt.east) coordinate (k2) {};
    \draw ([yshift=-1.5mm] gain.west) coordinate (k1) {};

    \begin{scope}[start chain]

        \chainin (tm);
        \chainin (tm_m)       [join=by dspline];
        \chainin (tm_l)       [join=by dspline];
        \chainin (tm_l_)       [join=by dspconn];
        \chainin (tm_m);
        \chainin (tm_r)       [join=by dspline];
        \chainin (tm_r_)       [join=by dspconn];

        \chainin (she);
        \chainin (she_m)       [join=by dspline];
        \chainin (she_l)       [join=by dspline];
        \chainin (she_l_)       [join=by dspconn];
        \chainin (she_m);
        \chainin (she_r)       [join=by dspline];
        \chainin (she_r_)       [join=by dspconn];

        \chainin (y_in_2);      
        \chainin (y_in_3)       [join=by dspconn];
        
        \chainin (ref);
        \chainin (M0_pre)       [join=by dspline];
        \chainin (M0)       [join=by dspconn];
    
        \chainin (gain);
        \chainin (filt)       [join=by dspconn];
        
        \chainin (filt);
        \chainin (p3)       [join=by dspline];
        \chainin (T2)       [join=by dspconn];
        \chainin (M1)       [join=by dspconn];

        \chainin (c4_3);
        \chainin (M0)       [join=by dspconn];
        \chainin (M1)       [join=by dspconn];
        \chainin (A1)       [join=by dspconn];

        \chainin (y);
        \chainin (A1) 	[join=by dspconn];
        \chainin (e) 	[join=by dspconn];

        \chainin (p2);
        \chainin (c3) 	[join=by dspline];
        \chainin (c4) 	[join=by dspline];
        \chainin (y_low1) 	[join=by dspconn];

        \chainin (c4_2) ;
        \chainin (y_low2) 	[join=by dspconn];
        
        \chainin (c5);
        \chainin (x_low1) 	[join=by dspconn];
        \chainin (x_in_1);
        \chainin (c5_2)     [join=by dspline];
        \chainin (x_low2) 	[join=by dspconn];
        \chainin (c5_3);
        \chainin (x_low3) 	[join=by dspconn];

        \chainin (p3);
        \chainin (T) 	    [join=by dspconn];
        \chainin (c7) 	    [join=by dspline];
        \chainin (gain) 	[join=by dspconn];
        
    \end{scope}
	
\end{tikzpicture}
    \end{minipage}
    \hfill
    \begin{minipage}{0.42\columnwidth}
        	\usetikzlibrary{dsp}
	\usetikzlibrary{chains}
	\usetikzlibrary{shapes.geometric}
	\usetikzlibrary{calc}

\begin{tikzpicture}[scale = 0.9]

\pgfarrowsdeclare{dsparrow}{dsparrow}
{
	\arrowsize=0.40pt
	\advance\arrowsize by .5\pgflinewidth
	\pgfarrowsleftextend{-4\arrowsize}
	\pgfarrowsrightextend{4\arrowsize}
}
{
	\arrowsize=0.40pt
	\advance\arrowsize by .5\pgflinewidth
	\pgfsetdash{}{0pt} 
	\pgfsetmiterjoin	 
	\pgfsetbuttcap		 
	\pgfpathmoveto{\pgfpoint{-4\arrowsize}{2.5\arrowsize}}
	\pgfpathlineto{\pgfpoint{4\arrowsize}{0pt}}
	\pgfpathlineto{\pgfpoint{-4\arrowsize}{-2.5\arrowsize}}
	\pgfpathclose
	\pgfusepathqfill
}

\tikzstyle{encdec}=[trapezium, trapezium angle=67.5, draw, inner ysep=5pt, outer sep=0pt,text width=1, minimum height=10, line width=.8pt, fill=white]

\pgfmathsetmacro{\toprow}{3.0}
\pgfmathsetmacro{\bottomrow}{1}
\pgfmathsetmacro{\middlerow}{(2.10}
\pgfmathsetmacro{\boxheight}{2}
\pgfmathsetmacro{\boxwidth}{5.5}
\pgfmathsetmacro{\hscale}{1.6}
\pgfmathsetmacro{\vscale}{1}

    \draw   (-2.55*\hscale, -0.1*\vscale) node    (backdrop)     [rectangle, rounded corners=0.1cm, minimum width=18 em, minimum height=18.5 em, fill=lightblue] {}

            (0.0*\hscale, -3.5*\vscale) coordinate (y) {}

            (-0.0*\hscale, -0.3*\vscale) coordinate (c5) {} 

            (-1.0*\hscale, -0.4*\vscale) coordinate (c3_2) {} 
            (-2.0*\hscale, -0.4*\vscale) coordinate (c4_2) {} 
            (-2.4*\hscale, -0.4*\vscale) coordinate (c5_2) {} 
            (-3.0*\hscale, -0.4*\vscale) coordinate (c3_3) {} 
            (-1.1*\hscale, 3.7*\vscale) coordinate (c5_3) {}
            (-4.0*\hscale, 3.88*\vscale) coordinate (c4_3) {} 

            (-3*\hscale, 3.7*\vscale) coordinate (y_in_2) {} 

            (-4.0*\hscale, -0.4*\vscale) node (x_hat_1) [dspnodefull] {}

            (-2*\hscale, +2.75*\vscale) node (she) [dspfilter, minimum width=2.2*\boxwidth em, rounded corners=0.1cm, minimum height=2.5 em, fill=red!20] {}
            (-2*\hscale, +2.75*\vscale) node (t1) [align=center, text width=2.4*\boxwidth em] {Distortion Model / \\ Kalman Gain} 
            
            (-1*\hscale, +2.25*\vscale) coordinate (she_r) {}
            (-3*\hscale, +2.25*\vscale) coordinate (she_l) {}
            (-1*\hscale, -1.3*\vscale) coordinate (she_r_) {}
            (-3*\hscale, +1.85*\vscale) coordinate (she_l_) {}

            (-3.0*\hscale, 1.75*\vscale) coordinate (M0_pre)
            (-4.0*\hscale, 1.85*\vscale) node  (M0In) [circle, inner sep= 4, fill=white]  {}
            (-4.0*\hscale, 1.85*\vscale) node (M0) [dspmixer] {}

            (-4*\hscale, 0.5*\vscale) node (buff) [dspfilter, minimum width=\boxwidth em, rounded corners=0.1cm, minimum height=2.5 em, fill=blue!10] {}
            (-4*\hscale, 0.5*\vscale) node (t1) [align=center, text width=\boxwidth em] {Ring \\ Buffer}

            (-2*\hscale, -1.3*\vscale) node (filt) [dspfilter, minimum width=\boxwidth em, rounded corners=0.1cm, minimum height=2.5 em, fill=blue!10] {}
            (-2*\hscale, -1.3*\vscale) node (t1) [align=center, text width=\boxwidth em] {Filter-State \\ Update} 

            (-4.0*\hscale, -2.5*\vscale) node  (M1In) [circle, inner sep= 4, fill=white]  {}
            (-4.0*\hscale, -2.5*\vscale) node (M1) [dspmixer] {} 
            (-4.0*\hscale, -3.5*\vscale) node  (AIn) [circle, inner sep= 4, fill=white]  {}
            (-4.0*\hscale, -3.5*\vscale) node (A1) [dspadder] {}

            (+0.1*\hscale, -0.4*\vscale) coordinate (c1) {} 
            (-4.0*\hscale, -0.4*\vscale) coordinate (c2) {} 

            (-2*\hscale, -2.5*\vscale) coordinate (p3) {} 
            (+0.0*\hscale, -2.5*\vscale) coordinate (c7) {} 

            (-0.5*\hscale, -3.5*\vscale) node (p2) {} 

            (-5.4*\hscale, -3.5*\vscale) coordinate (e) {}

            (-2.6*\hscale, -2.5*\vscale) node (T2) [dspfilter, minimum width=1.5 em, minimum height=1.5 em, fill=white] {T}
            
            ;

    \draw ([xshift=5mm, yshift=+1.0mm] M0_pre.east) node (Y_t) {${M}_{\ell,k}$};

    \draw ([xshift=0mm, yshift=+2.0mm] c5_3.east) node (X_t) {${\mathbf{X}}_{\ell,k}$};
    \draw ([xshift=+6mm, yshift=+3.0mm] p2.east) node (Y_t) {${Y}_{\ell,k}$};

    \draw ([xshift=0mm, yshift=+2.0mm] y_in_2.east) node (X_t) {${Y}_{\ell,k}$};

    \draw ([xshift=0mm, yshift=+3.0mm] c4_2.east) node (Y_t) {${E}_{\ell,k}$};
    \draw ([xshift=5mm, yshift=+0.0mm] c4_3.east) node (Y_t) {${Y}_{\ell,k}$};

    \draw ([xshift=+5.0mm, yshift=-3.5mm] filt.east) node (Y_t) {$\mathbf{K}_{\ell,k}$};

    \draw ([xshift=+7mm, yshift=-4.5mm] M0.west) node (X_t) {$\widehat{{X}}_{\ell,k}$};

    \draw ([xshift=+7mm, yshift=+16.5mm] M1.west) node (X_t) {$\widehat{\mathbf{X}}_{\ell,k}$};
    
    \draw ([xshift=-0mm, yshift=-3.5mm] p3.east) node (H_t) {$\widehat{\mathbf{H}}_{\ell,k}$};
    \draw ([xshift=-21mm, yshift=+3.5mm] p3.east) node (H_t) {$\widehat{\mathbf{H}}_{\ell-1,k}$};
    \draw ([xshift=+3.0mm, yshift=+5.0mm] A1.east) node (D_t) {$\widehat{D}_{\ell,k}$};

    \draw ([xshift=-5mm, yshift=+3.0mm] A1.east) node (min) {\huge -};
    \draw ([xshift=+5.5mm, yshift=+3mm] e.north) node (E_t) {${E}_{\ell,k}$};

    \draw ([yshift=+7.1mm] M1) coordinate (xhat_low) {};

    \draw ([yshift=-5.1mm] c5) coordinate (x_low1) {};
    \draw ([yshift=-4.1mm] c5_2) coordinate (x_low2) {};
    \draw ([yshift=-4.1mm] c4_2) coordinate (y_low2) {}; 
    \draw ([yshift=-4.51mm] c5_3) coordinate (x_low3) {};
    \draw ([yshift=-4.1mm] c4_3) coordinate (y_low3) {}; 

    \draw ([yshift=-4.5mm] y_in_2) coordinate (y_in_3) {}; 

    \draw ([yshift=1.5mm] filt.west) coordinate (a2) {};
    \draw ([xshift=2.5mm, yshift=1.5mm] filt.east) coordinate (a3) {};

    \draw ([yshift=-1.5mm] filt.east) coordinate (k2) {};

    \begin{scope}[start chain]

        \chainin (she_l);
        \chainin (she_l_)       [join=by dspline];
        \chainin (M0)           [join=by dspconn];
        \chainin (she_r);
        \chainin (she_r_)       [join=by dspline];
        \chainin (filt)         [join=by dspconn];

        \chainin (y_in_2);       
        \chainin (y_in_3)       [join=by dspconn];
        
        \chainin (filt);
        \chainin (p3)       [join=by dspline];
        \chainin (T2)       [join=by dspconn];
        \chainin (M1)       [join=by dspconn];

        \chainin (c4_3);
        \chainin (M0)       [join=by dspconn];
        \chainin (buff)     [join=by dspconn];
        \chainin (M1)       [join=by dspconn];
        \chainin (A1)       [join=by dspconn];

        \chainin (y);
        \chainin (A1) 	[join=by dspconn];
        \chainin (e) 	[join=by dspconn];

        \chainin (c4_2) ;
        \chainin (y_low2) 	[join=by dspconn];
        
        \chainin (x_hat_1);
        \chainin (c5_2)     [join=by dspline];
        \chainin (x_low2) 	[join=by dspconn];
        \chainin (c5_3);
        \chainin (x_low3) 	[join=by dspconn];
        
    \end{scope}
	
\end{tikzpicture}
    \end{minipage}
    \vspace{-0.2mm}
	\caption{Details of the Kalman algorithm block as used in the {\tt NeuralKalman}~\cite{Zhang2023} (left) and {\tt DeepAdaptive}~\cite{Zhang2022d} (right) solutions. Red blocks are realized by a DNN, while the filter-state update of {\tt NeuralKalman} is partially DNN-based.}
	\label{fig:NKal}
 \end{figure}

The {\tt NeuralKalman} model~\cite{Zhang2023}, as shown in Table\,\ref{tab:diff} and in Fig.~\ref{fig:NKal}, uses a DNN to estimate the \textit{dynamic} state transition factor $A_{\ell}^\mathrm{{DNN}}({\mathbf{X}}_{\ell},{\mathbf{Y}}_{\ell})$ valid for all frequency bins $k$ in frame $\ell$. It is noteworthy that while the model as proposed by the authors operates on a single-tap filter estimation, the shared feature encoder for the distortion and transition model uses several past frames to compute correlation of frequency bins over time, frequency correlation, channel correlation, and normalized log-power spectrum as input features, which are then processed by a 4-layer LSTM block. The transition model itself only consists of a linear layer with a sigmoidal activation function. 

\subsection{DNN-Based Filter State Update}
\label{ssec:Update_DNN}

While replacing the Kalman gain and the state transition factor can improve (re)convergence speed, those approaches would not be of use in solving issues arising from nonlinear distortions. Direct manipulation of the final filter state, e.g., by postprocessing of its classical calculation, can better enable the system to model nonlinear distortions.

The \texttt{NeuralKalman} approach~\cite{Zhang2023} investigated in this article employs an LSTM-based DNN as a postprocessing function to the filter state $\widehat{{H}}^\mathrm{{SP}}_{\ell,k}$ already updated according to (\ref{eq:H_FDKF}) of the Kalman filter framework,
\begin{align}
\label{eq:state-update-neuralkalman}
    \widehat{{H}}^\mathrm{{DNN}}_{\ell,k}({{X}}_{\ell,k},{{E}}_{\ell,k}, {{K}}^\mathrm{SP}_{\ell,k}) = \widehat{{H}}^\mathrm{{DNN}}_{\ell,k}\Big(\widehat{{H}}^\mathrm{{SP}}_{\ell,k}\big({{X}}_{\ell,k},{{E}}_{\ell,k}, {{K}}^\mathrm{SP}_{\ell,k}\big)\Big)\;, 
\end{align}
whereby the utilization of a DNN-based Kalman gain would be yet another option.

Beyond the scope of this article, the aforementioned \texttt{Meta-AF}~\cite{MetaAF2023} replaces both Kalman gain and filter-state update blocks by a DNN similar to those of {\tt NKF} and {\tt DLAC-Kalman}, obtaining a learned filter-state update rule 
\begin{equation}
\label{eq:metaaf-state-update}
    \widehat{{\VEC{H}}}_{\ell,k} = \widehat{{\VEC{H}}}_{\ell-1,k} + \VECG{\Delta}^\mathrm{{DNN}}_{\ell,k}({E}_{\ell,k},{\VEC{X}}_{\ell,k},{Y}_{\ell,k}, \VECG{\nabla}_{\ell,k})
\end{equation}
directly from training data instead of a hand-crafted algorithm. As such, \texttt{Meta-AF} does not bear a notion of Kalman or neural Kalman filtering, however, note the classical gradient $\boldsymbol{\nabla}_{\ell,k} = {X}^*_{\ell,k} {E}_{\ell,k}$ among its inputs.

\subsection{DNN-Based Distortion Modeling}
\label{ssec:Distortion_DNN}

Another, more straightforward way of dealing with nonlinear distortions caused by the loudspeaker is to estimate and apply these distortions directly to the reference signal ${\mathbf{X}}_{\ell,k}$ and, in this way, to revert to a subsequent linear FDKF model acting on the predistorted reference. 

Both {\tt NeuralKalman}~\cite{Zhang2023} and {\tt DeepAdaptive}~\cite{Zhang2022d} employ a similar approach to modelling distortions: A DNN---which notably is (partly) shared with other estimations in both cases (see Fig.~\ref{fig:NKal})---computes a mask ${M}_{\ell,k}$ ({\tt DeepAdaptive}) or time-frequency complex-valued deep filtering kernel $\MAT{M}_{\ell,k}\in\C^{3\times 3}$~\cite{Mack2020} ({\tt NeuralKalman}) which is applied to the \textit{microphone} signal to create the nonlinear reference signal estimate ${\widehat{X}}_{\ell,k}$. For {\tt NeuralKalman}, we employ a $3 \times 3$ causal kernel to harmonize algorithmic delay and temporal context for all approaches under investigation (see p.\ \pageref{ssec:framework}).
Both DNNs use the microphone and reference signals as input, but for the {\tt NeuralKalman} model, multiple taps of both signals are used in the feature encoder.
Another difference between the two approaches can be found in the application of ${\widehat{X}}_{\ell,k}$. While both models compute the echo estimate according to \mbox{$\widehat{D}_{\ell,k} = \widehat{\VEC{X}}_{\ell,k}^T \cdot \widehat{{\VEC{H}}}_{\ell-1,k}$}, only the {\tt DeepAdaptive} model uses ${\widehat{X}}_{\ell,k}$ to replace the original reference signal ${{X}}_{\ell,k}$ in the filter-state update.

Using the microphone signal as basis for the estimate ${\widehat{X}}_{\ell,k}$ may have significant implications on the entire model behavior. Let's for {mathematical} simplicity assume a mask or $1 \!\times\! 1$ filter kernel ${M}_{\ell,k}$ generated from either method. Hence, the nonlinear reference signal estimate would calculate as ${\widehat{X}}_{\ell,k} = {M}_{\ell,k} \cdot {Y}_{\ell,k}$. With \eqref{eq:H_FDKF}, we can infer
\begin{equation}
    {{E}}_{\ell,k} = {Y}_{\ell,k} - {M}_{\ell,k} \cdot {Y}_{\ell,k} \cdot \widehat{{{H}}}_{\ell-1,k} = {Y}_{\ell,k} \cdot {G}_{\ell,k},
\end{equation}
with gain ${G}_{\ell,k} = 1 - {M}_{\ell,k} \cdot \widehat{{{H}}}_{\ell-1,k}$ applicable to our microphone signal, defining our entire neural Kalman filtering. Consequently, the approach {may fall} into the class of mask-based echo suppressors instead of subtractive AEC (cf. Fig.~\ref{fig:systemlvl}). This has implications on the expected model behaviour, as echo suppressors are often more efficient in echo reduction at the cost of more distortion to the near-end speech signal. 
Furthermore, since no intermediate losses enforce the focus of the distortion models ${M}_{\ell,k}$ on their intended task, it is difficult to control which role ${M}_{\ell,k}$ and $\widehat{{{H}}}_{\ell-1,k}$ will finally play in the trained model. In an extreme case, {\tt NeuralKalman} may, for instance, learn $\widehat{{{H}}}_{\ell-1,k}=1$ and solely rely on ${M}_{\ell,k}$ for estimation of ${G}_{\ell,k}$, in that case even detaching the Kalman filter loop from the system.

\subsection{State as Conceptual Bridge Between Model-Based and DNN-Based Paradigms}

While the article has been concerned so far with the status of merging the strongly model-based Kalman filter framework with the data-driven neural network approach, the respective terminologies according to different sources of origin may somewhat puzzle the understanding of currently proposed systems. The idea of this subsection therefore is to unite the seemingly orthogonal concepts of trainable (then fixed) weights of a neural network and the adaptive weights of a model-based adaptive filter, for instance, the Kalman filter.

One angle of view in this respect might be the understanding that neural networks, rich in trained weights and with sufficiently diverse input features, bear the subtlety for delivering very specific, for instance, time-varying (here in the sense of ''adaptive'') predictions for the specific problem at hand. In this case, the DNN would just need to be sufficiently large to mimic or even improve, for instance, the behavior of a model-based KF. The potentially unreasonable size of the network in this ideology might be efficiently reduced by employing useful model-based architectures as a baseline and by compensating its limitations with additional DNN blocks.

Our take here is a bit more tangible in that the concept of model state overarches the model-based adaptive-filter world and the data-driven neural-network world. The understanding of model state is compatible in both worlds and greatly unites and eventually enlarges the available system options. In the adaptive-filter world, if we consider the adaptive weights as a system state to be updated, its architectural counterpart in the DNN world is simply the state of a recurrent unit. Conversely, if a recurrent state update is accomplished by a rich set of trained weights acting on the current input signal, its model-based counterpart is a typically sophisticated mathematical function to process the same input. We can therefore easily combine the computational paradigms of both worlds.

In our article, we have defined the neural Kalman filters such that the state update equation (\ref{eq:H_FDKF}) of FDKF is intrinsic to the systems under investigation. The diversity of system options is then created by the specification of different architectures for actual computation of the state update, for instance, a DNN-based nonlinear postprocessing of the resulting filter states according to (\ref{eq:state-update-neuralkalman}) by \texttt{NeuralKalman}, or a DNN-based Kalman gain estimation according to the $\VEC{K}^{\mathrm{DNN}}_{\ell,k}$ cases in our portfolio of Table~\ref{tab:diff}, hence including the \texttt{DeepAdaptive} AEC.

\section{Comparative Evaluation of Methods and Discussion}
\label{sec:Comparison}

In this section, we will provide a comparative experimental analysis and discussion of {\tt FDKF} (realized as described on pp.\ \pageref{sec:FDKF}ff.) and its hybrid variants with DNN support. Various test conditions with single-talk, double-talk, real-world RIRs, echo path changes and nonlinear loudspeaker distortions different from training will be evaluated. Beyond ERLE, we report AECMOS~\cite{Purin2021},
PESQ~\cite{ITU_P862.2_Corr2}, and STOI \cite{Taal2011} for delivering a comprehensive picture of performance and limitations of the various approaches. All methods operate on data sampled at $16$ kHz {(wideband speech)}. Any databases of originally higher sampling rate are down-sampled before use in data generation.

\subsection{Common Processing Framework}
\label{ssec:framework}

All compared methods \cite{Zhang2022d,Zhang2023,Yang2023,Haubner2024} have been implemented for {the aforementioned} $16$ kHz sampling rate with the same algorithmic delay. Overlap-save (OLS) as used in the {\tt FDKF} is typically employed for linear convolution in system modelling, where $K=N+R-1$ holds, with $R$ being the number of new samples in a DFT frame (i.e., frame shift), and $N$ being the maximum modelled impulse response length. The OLS algorithmic delay is thus equal to the frame shift $R$. In contrast, overlap-add (OLA) methods in speech enhancement are often performing circular convolution (i.e., no zero padding), and exhibit algorithmic delay equal to the DFT size $K$. Accordingly, we choose $K^\mathrm{OLA}=R^\mathrm{OLS}=512$ {to constrain to the same delay of OLS- and OLA-based systems}, whereby the employed 75\% overlap in OLA means \mbox{$R^\mathrm{OLA}=K^\mathrm{OLA}/4$}. {For fair comparison of single-tap vs.\ multi-tap methods we further define an \textit{effective reference input length} $M= K + (L-1)\cdot R$, i.e., the total amount of unique reference samples utilized for a single filter state update, which applies to both OLA and OLS. This temporal context is aligned to the same size for all models of this article. Specifically, the multi-tap OLAs employ} a DFT size of $K=512$ and $L=8$ frame taps with $R=128$, while single-tap ($L=1$) OLS methods use a DFT size of $K=1408$ with $R=512$. OLS multi-tap methods use a DFT size of $K=896$ and $L=2$ frame taps, also with $R=512$. These settings are unique for this article to allow a fair comparison and may deviate for the various methods from authors' original choices.

\subsection{Training Description}
\label{ssec:training}

All hybrid models are trained from scratch in our \texttt{PyTorch2}~\cite{Paszke2019} framework, using a {\tt GTX 1080 Ti} {\rm GPU}. To generate speech utterances for the training dataset $\mathcal{D}_{\mathrm{train}}$, we use speech files from the publicly available CSTR-VCTK corpus~\cite{VCTK}, with each speaker serving as both far-end and near-end speaker in separate cases, thereby avoiding overfitting to specific voices. The scaled error function (SEF)~\cite{Klippel2006}, defined as 
\begin{equation}
    x'(n) = f_\mathrm{SEF}(x(n)) = \int_{0}^{x(n)} \exp\left({\frac{z^2}{2\alpha^2}}\right) dz,
    \label{eq:SEF}
\end{equation}
is employed to simulate loudspeaker nonlinearities (NL), distorting the FE reference signal $x(n)$, with $\alpha$ randomly chosen from the set $\{0.5, 1, 10, 999\}$ for each audio file. The loudspeaker signal $x'(n)$ is then convolved with an RIR of a reverberation time randomly sampled from a continuous range according to RT$60 \in [50, 600]$\,ms. In training, we generate these RIRs by modulating a white Gaussian noise signal with an exponential decay \cite{Jung2014}. The NE signal-to-echo ratio (SER) is randomly sampled from a continuous range of $[-12.4,22.4]$\,dB for each audio file.
To make the models more robust against background noise, we add random cuts from the publicly available DEMAND~\cite{Thiemann2013} and QUT-NOISE~\cite{QUT} databases at a signal-to-noise ratio (SNR) in the range of $[-2.4, 32.4]$\,dB to $90$\% of microphone signals. In total, $9500$ audio files of $10$\,s length are generated for training, with $1000$ being split off into a subset $\mathcal{D}_{\mathrm{dev}}$, which is exclusively used for learning rate control and early stopping criteria. To diversify our training data, the components $s(n)$, $n(n)$, and $d(n)$ of these audio files (except $\mathcal{D}_{\mathrm{dev}}$) are randomly shuffled with new SER/SNR values between each epoch.

All models are trained using the Adam optimizer~\cite{Kingma2015} in its standard configuration, and a logarithmic MSE loss function solely aiming at acoustic echo cancellation,
\begin{align}
    {J}^{\mathrm{logMSE}} & = 10\!\cdot\!\log \Big(\sum_{n \in \mathcal{N}} \big(e(n)-s(n)-n(n)\big)^2  \Big),
    \label{eq:tlmse}
\end{align}
computed over the entire time sequence $(n \in \mathcal{N} \subset \mathbb{N}_0)$. 
The training batch size is set to 8, with a backpropagation-through-time (BPTT) unrolling depth of $100 \cdot R^\mathrm{OLA} = 12800$ samples. The training process starts at a learning rate (LR) of $10^{-3}$, which is halved after 4 consecutive epochs without loss improvement on $\mathcal{D}_{\mathrm{dev}}$. The training is stopped after 100 epochs, or if the LR falls below $10^{-5}$, or if the loss on $\mathcal{D}_{\mathrm{dev}}$ does not improve for 10 consecutive epochs.

\subsection{Evaluation Data, Settings, and Metrics}
\label{ssec:Settings}

For our test setup, we create a dataset distinct from the training material. Speakers for FE and NE are taken from the TIMIT speech corpus~\cite{TIMIT}. Instead of modelling the RIRs synthetically, we use real-world recordings from the Aachen Impulse Response database \cite{Jeub2010}, which we only modify to exclude initial delay in the echo path. Background noise is sampled from the publicly available ETSI noise database~\cite{ETSI2008}, specifically choosing environments so far unseen in training noise.
The SER is randomly chosen from $\{-9, -6, ..., 9\}$\,dB, while the SNR is fixed at $20$\,dB, thereby representing a moderate (unsuppressed) background noise that does not have a significant effect on the employed metrics while still imposing a bit of a challenge on the evaluated models. Note that the broad range of SER values naturally leads to large standard deviations in most metrics. Accordingly, we do not report standard deviations.

We test all models under various conditions for a precise understanding of their capabilities and limitations. For this, we create three distinct test sets: The basic test set $\mathcal{D}_{\mathrm{test}}$ contains \textit{speech} FE excitation and no nonlinear loudspeaker distortion model. The derivative test set $\mathcal{D}_{\mathrm{test}}^\mathrm{WGN}$ substitutes the FE speaker with \textit{white Gaussian noise} (WGN) excitation, while test set $\mathcal{D}_{\mathrm{test}}^\mathrm{NL}$ deviates from $\mathcal{D}_{\mathrm{test}}$ by a \textit{harsh nonlinear distortion} through the memoryless sigmoidal function \cite{Zhang18g} (training/test mismatch). We generate 60 files per test set, each consisting of three 8\,s long consecutive sections with varying speaker activity: firstly, single-talk far-end (STFE), secondly, single-talk near-end (STNE), and finally double-talk (DT). In the following, results are usually reported in the context of a single section, while in the case of DT evaluation, the preceding sections are only employed for initial convergence towards steady-state performance. In some test conditions, RIR switches may occur in the middle (at the 4\,s mark) of a section.

There are two main aspects to evaluate in our models: a model's capability to remove echo, and its capability of maintaining an undistorted and intelligible NE speech signal. Our major evaluation metric for echo cancellation effectiveness here is the echo-return loss enhancement (ERLE) after \cite{Vary2006},
\begin{equation}
    \mathrm{ERLE}(n)=10\cdot\mathrm{log}_{10}\left(\frac{\EV{d^2(n)}}{\EV{\big(d(n)-\widehat{d}(n)\big)^2}}\right) \quad \text{in dB},
    \label{eq:ERLE}
\end{equation}
with echo estimate $\widehat{d}(n)=y(n)-e(n)$, the expectation operator $\EV{\cdot}$ approximated by a first-order IIR smoothing filter with impulse response $g(n)=\alpha^n, n\in\{0,1,2,...\}$, and coefficient \mbox{$\alpha=0.99$}.
To further evaluate the effectiveness of a model to cancel the echo, we use the AECMOS metric~\cite{Purin2021}, which aims to estimate the MOS scores of a subjective listening test regarding echo reduction (DT\,Echo) and NE speech quality (DT\,Other). For additional evaluation of the NE speech preservation, we also use the Perceptual Evaluation of Speech Quality (PESQ) metric~\cite{ITU_P862.2_Corr2}. The  Short-Time Objective Intelligibility (STOI) metric~\cite{Taal2011} is also employed for a more focused observation of NE speech intelligibility. Furthermore, we use the Levenshtein Phoneme Similarity (LPS) metric\footnote{Pirklbauer et al. \cite{Pirklbauer2023} define the Levenshtein Phoneme Distance (LPD), whereby $\mathrm{LPS}=1-\mathrm{LPD}$.}~\cite{Pirklbauer2023} to measure phonetic fidelity in a language-independent way. As it is highly correlated with automatic speech recognition (ASR) results as well as overall speech enhancement performance, it is well suited to rate performance beyond the mere AEC application (e.g., as preprocessing for ASR). PESQ, STOI, and LPS are evaluated with the clean NE speech signal $s(n)$ as reference signal. Note that all of our metrics, the {\tt FDKF}, and many useful Python training and evaluation scripts are available under \mbox{\url{https://github.com/ifnspaml/EC-Evaluation-Toolbox}}.

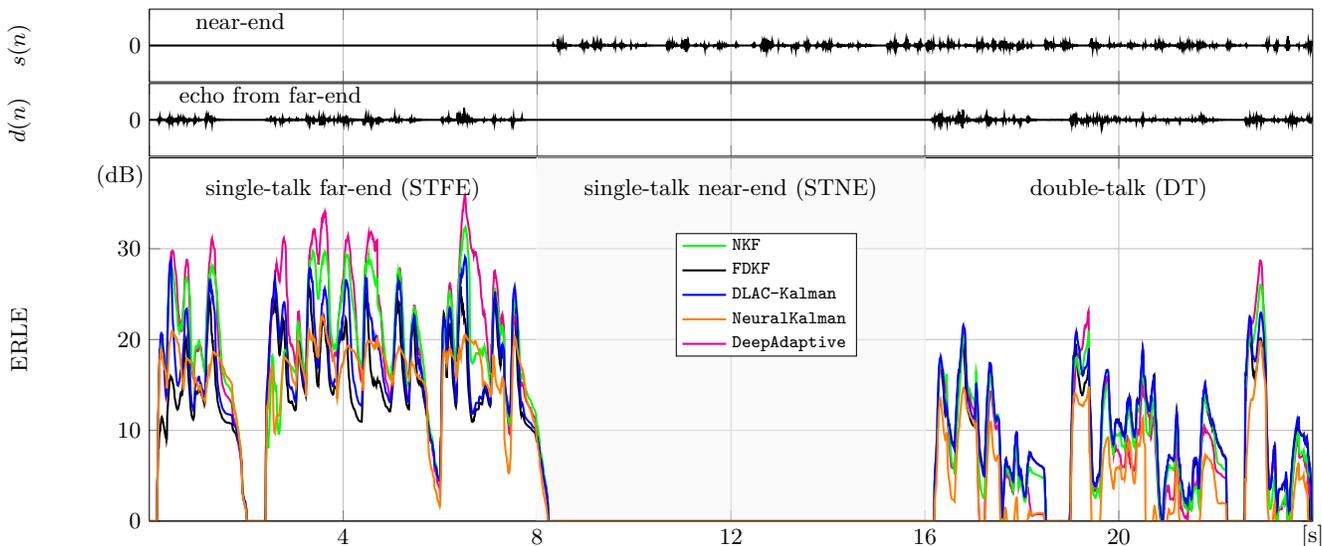
\begin{figure*}
    \centering
    \vspace{0mm} \hspace{-7mm}
    	\small
    \newlength\figureheightFS
    \newlength\figurewidthFS
    \setlength\figureheightFS{1.0cm}
    \setlength\figurewidthFS{16cm}
    \newlength\figureheightF
    \newlength\figurewidthF
    \setlength\figureheightF{5.0cm}
    \setlength\figurewidthF{16cm}

    \definecolor{mycolor1}{rgb}{0.749019622802734,0,0.749019622802734}%
    \begin{tikzpicture}
        \begin{axis}[%
            width=\figurewidthFS,
            height=\figureheightFS,
            scale only axis,
            name=NE plot,
            xmin=0,
            xmax=384000,
            xtick={0,64000,128000,192000,256000,320000},
            xticklabels={},
            xmajorgrids,
            scaled ticks=false,
            ymin=-1,
            ymax=1,
            ytick={0},
            ylabel={$s(n)$},
            label style={at={(-0.035,.5)}},
            ymajorgrids,
            legend style={at={(0.993,0.0125)},anchor=south east,draw=black,legend cell align=left, fill=white, row sep=0, inner sep=0, inner xsep=2pt,legend columns=1, font={\scriptsize}}
            ]

                \draw [black, dashed] (axis cs: 502000,-1.0) -- (axis cs: 502000,1.0);
		
                \addplot [ thick, no marks, black] file {fig/plots_20dB/single/Kalman_long8_eqD_SER0_SNR0_BB_s.txt};

                \draw (axis cs: 30000, 0.68) node (text) {near-end};
                
        \end{axis}


        \begin{axis}[%
            at={(NE plot.below south west)},
            yshift=-0.8cm,
            name = FE plot,
            width=\figurewidthFS,
            height=\figureheightFS,
            scale only axis,
            xmin=0,
            xmax=384000,
            xtick={0,64000,128000,192000,256000,320000},
            xticklabels={},
            xmajorgrids,
            scaled ticks=false,
            ymin=-1,
            ymax=1,
            ytick={0},
            ylabel={$d(n)$},
            label style={at={(axis description cs:-0.035,.5)}},
            ymajorgrids,
            legend style={at={(0.993,0.0125)},anchor=south east,draw=black,legend cell align=left, fill=white, row sep=0, inner sep=0, inner xsep=2pt,legend columns=1, font={\scriptsize}}
            ]

                \draw [black, dashed] (axis cs: 502000,-1.0) -- (axis cs: 502000,1.0);
		
                \addplot [ thick, no marks, black] file {fig/plots_20dB/single/Kalman_long8_eqD_SER0_SNR0_BB_d.txt};

                \draw (axis cs: 40000, 0.68) node (text) {echo from far-end};
                
        \end{axis}


        \begin{axis}[%
            at={(FE plot.below south west)},
            anchor=north west,
            yshift=+0.2cm,
            width=\figurewidthF,
            height=\figureheightF,
            scale only axis,
            xmin=0,
            xmax=384000,
            xtick={64000,128000,192000,256000,320000},
            xticklabels={4,8,12,16,20},
            xmajorgrids,
            xminorgrids,
            scaled ticks=false,
            ymin=0,
            ymax=40,
            ytick={-10,0,10,20,30},
            ylabel={ERLE},
            label style={at={(axis description cs:-0.035,.5)}},
            ymajorgrids,
            legend style={at={(0.993,0.0125)},anchor=south east,draw=black,legend cell align=left, fill=white, row sep=0, inner sep=0, inner xsep=2pt,legend columns=1, font={\scriptsize}}
            ]

                \draw [fill=gray!7, draw=white!0, fill opacity=0.5] (axis cs: 128000,-30.0) rectangle (axis cs: 256000,45.0);

                \addplot[ thick, no marks, black] file {fig/plots_20dB/single/Kalman_long8_eqD_SER0_SNR0_BB_EoT.txt};

                \addplot[ thick, no marks, magenta] file {fig/plots_20dB/single/DeepAdaptive_AEC_model_8_LogMSE_LogMSEMS_ERNST_cluster_ern_CL_VCTKem_16_1080_1_full_fin_SER0_SNR0_BB_EoT.txt};
                \addplot[ thick, no marks, green] file {fig/plots_20dB/single/NKF_AEC_model_8_LogMSE_LogMSEMS_ERNST_cluster_ern_CL_VCTKem_16_1080_1_full_stepped_SER0_SNR0_BB_EoT.txt};

                \addplot[ thick, no marks, blue] file {fig/plots_20dB/single/DLAC_AEC_model_final_v2_noNorm_LogMSE_LogMSEMS_ERNST_cluster_ern_CL_VCTKem_16_1080_1_full_SER0_SNR0_BB_EoT.txt};

                \addplot[ thick, no marks, orange] file {fig/plots_20dB/single/NeuralKalman_AEC_model_1_k_LogMSE_LogMSEMS_ERNST_cluster_ern_CL_VCTKem_16_1080_1_full_f3_shortSeq_SER0_SNR0_BB_EoT.txt};

                \draw (axis cs: 64000,36.5) node (t) [text width=4cm, align=center] {single-talk far-end (STFE)};

                \draw (axis cs: 192000,36.5) node (t) [text width=5cm, align=center] {single-talk near-end (STNE)};

                \draw (axis cs: 320000,36.5) node (t) [text width=3cm, align=center] {double-talk (DT)};

                \draw [black] (axis cs: 0,0.0) -- (axis cs: 640000,0.0);

        \end{axis}

        \begin{axis}[%
            at={(FE plot.below south west)},
            anchor=north west,
            yshift=+0.2cm,
            hide axis,
            width=\figurewidthF,
            height=\figureheightF,
            xmin=0,
            xmax=64000,
            ymin=0,
            ymax=22,
            legend style={at={(0.675,0.2)},anchor=south east, draw=black,legend cell align=left, fill=white, row sep=0, inner sep=0, inner xsep=2pt,legend columns=1, font={\scriptsize}}
            ]

            \addlegendimage{thick, no marks, green}
            \addlegendentry{{\tt NKF}}
            \addlegendimage{thick, no marks, black}
            \addlegendentry{{\tt FDKF}}
            \addlegendimage{thick, no marks, blue}
            \addlegendentry{{\tt DLAC-Kalman}}
            \addlegendimage{thick, no marks, orange}
            \addlegendentry{{\tt NeuralKalman}}
            \addlegendimage{thick, no marks, magenta}
            \addlegendentry{{\tt DeepAdaptive}}
            
        \end{axis}

    \node (s1) at (-0.4,-1.3){(dB)};
    \node (s1) at (16,-6.25){[s]};
    \end{tikzpicture}%
 
    \vspace{-2.5mm}
    \caption{Model performance for an example file from test set $\mathcal{D}_{\mathrm{test}}$, represented by ERLE 
    over time (bottom panel). Near-end speech $s(n)$ (top panel) and far-end echo $d(n)$ (center panel) are mixed at $0$\,dB SER.}
    \label{fig:example}
    \vspace{-3mm}
\end{figure*}

\subsection{Experiments: Convergence, Steady-State Performance, Reconvergence}
\label{ssec:Reconvergence}

The ability to quickly (re)converge towards a stable and potent filter state is an important characteristics of an AEC system. As it will be clear from the upcoming evaluation results, classical methods such as the {\tt FDKF} possess a rather pronounced reconvergence phase, making them vulnerable to quick echo path changes. 
As discussed on pp.\ \pageref{sec:Hybrid}ff., the presented hybrid models propose various solutions to increase convergence speed.

Fig.~\ref{fig:example} shows an example file from our data setup taken from $\mathcal{D}_{\mathrm{test}}$ (SER = $0$\,dB), which contains speech as FE excitation without nonlinear distortion. The ERLE scores  \eqref{eq:ERLE}
of all investigated models are plotted over time, alongside the NE speech $s(n)$ and echo from the FE $d(n)$ for all three conditions STFE, STNE, DT. In the initial STFE segment, we observe that all neural Kalman filters ramp up to a higher ERLE than the too smoothly updating \texttt{FDKF}. The \texttt{DeepAdaptive} model is clearly on top of the contenders reaching at times an ERLE of 30 dB or higher. Among the hybrid approaches, \texttt{NeuralKalman} reveals lowest ERLE, however, with a smaller drop of ERLE during low-power FE signals.
In the DT condition, the overall picture is similar but shows an about 10 dB lower ERLE. The former strength of \texttt{NeuralKalman} on low-power signals now converts to the opposite, as now a NE signal disturbs the echo path estimation process. Also the formerly strong \texttt{Deep Adaptive} method reveals ERLE plunges in high NE signal-to-echo (SER) segments. Not all hybrid methods show such STFE/DT trade-off: \texttt{NKF} and \texttt{DLAC-Kalman} have quite balanced performance in both conditions; the former being stronger in STFE and the latter being top-performing in DT. Looking into Tab.\ \ref{tab:diff}, we identify these two methods as the ones which merely replace the Kalman gain computation by a DNN. \texttt{NeuralKalman} and \texttt{DeepAdaptive} employ a DNN distortion model that---as has been discussed in detail on p. \pageref{ssec:Distortion_DNN}---supposedly changes their overall behaviour into that of mask-based echo suppressors, which expectedly have an advantage in the STFE condition.

Fig.~\ref{fig:over_time} displays our models' convergence behaviour by focusing on the STFE sections of $\mathcal{D}_{\mathrm{test}}$, with an RIR switch introduced after $4$\,s, to observe reconvergence as well. Also here, no nonlinearity has been simulated, i.e., $x'(n) = x(n)$. The ERLE scores are now averaged over the STFE sections of $8$\,s length of the entire test set $\mathcal{D}_{\mathrm{test}}$. The FE excitation is speech. This figure reconfirms the strength of neural Kalman filters when it comes to ERLE. In agreement with Fig.~\ref{fig:example}, \texttt{Deep Adaptive} in Fig.~\ref{fig:over_time} most dynamically ramps up highest and even ''reconverges'' rapidly. The latter observation also holds for \texttt{NeuralKalman}. 
This behaviour can once again be attributed to the echo suppression behaviour of both models (see p.\ \pageref{ssec:Distortion_DNN}), which allows for strong far-end single-talk performance and fast ''reconvergence''. Masking, which just needs to suppress the most dominant component in this condition, can be more aggressive compared to echo cancellation, the latter still being required to accurately estimate the room impulse response.
All other methods in Fig.~\ref{fig:over_time} show a performance drop at the time of the RIR switch (echo path change). The \texttt{FDKF} shows a quite poor temporal convergence and reconvergence performance. This slowness has been reported and explained earlier \cite{yang2017frequency}. Interestingly, the mean performance of {\tt NeuralKalman} seems to diverge from observations made in Fig.~\ref{fig:example}. Further inspection revealed that this method shows especially high echo suppression for low SER levels.

\begin{figure}[t!]
    \centering
    	\small
    \newlength\figureheightD
    \newlength\figurewidthD
    \setlength\figureheightD{3cm}
    \setlength\figurewidthD{14cm}

    \definecolor{mycolor1}{rgb}{0.749019622802734,0,0.749019622802734}%
    \begin{tikzpicture}
        \begin{axis}[%
            width=\figurewidthD,
            height=\figureheightD,
            scale only axis,
            xmin=800,
            xmax=126000,
            xtick={0,32000,64000,96000,128000},
            xticklabels={0,2,4,6},
            xmajorgrids,
            xminorgrids,
            minor x tick num=1,
            scaled ticks=false,
            ymin=0,
            ymax=27,
            ytick={-10,0,10,20},
            ylabel={ERLE},
            ymajorgrids,
            legend style={at={(0.993,0.0125)},anchor=south east,draw=black,legend cell align=left, fill=white, row sep=0, inner sep=0, inner xsep=2pt,legend columns=1, font={\scriptsize}}
            ]

                \addplot[ thick, no marks, black] file {fig/plots_20dB/STFE/Kalman_long8_eqD_CS_SER0_SNR0_BB_EoT.txt};

                \addplot[ thick, no marks, magenta] file {fig/plots_20dB/STFE/DeepAdaptive_AEC_model_8_LogMSE_LogMSEMS_ERNST_cluster_ern_CL_VCTKem_16_1080_1_full_fin_CS_SER0_SNR0_BB_EoT.txt};
                \addplot[ thick, no marks, green] file {fig/plots_20dB/STFE/NKF_AEC_model_8_LogMSE_LogMSEMS_ERNST_cluster_ern_CL_VCTKem_16_1080_1_full_stepped_CS_SER0_SNR0_BB_EoT.txt};

                \addplot[ thick, no marks, blue] file {fig/plots_20dB/STFE/DLAC_AEC_model_final_v2_noNorm_LogMSE_LogMSEMS_ERNST_cluster_ern_CL_VCTKem_16_1080_1_full_CS_SER0_SNR0_BB_EoT.txt};

                \addplot[ thick, no marks, orange] file {fig/plots_20dB/STFE/NeuralKalman_AEC_model_1_k_LogMSE_LogMSEMS_ERNST_cluster_ern_CL_VCTKem_16_1080_1_full_shortSeq_CS_SER0_SNR0_BB_EoT.txt};

                \draw [densely dashed, black] (axis cs: 64000,-30.0) -- (axis cs: 64000,45.0);

                \draw [black, ->,>={stealth}] (axis cs: 58000,4.2) -- (axis cs: 63000,7.0);

                \draw (axis cs: 52000,3.5) node (t) [text width=2cm] {RIR switch};

                \draw [black] (axis cs: 0,0.0) -- (axis cs: 640000,0.0);

        \end{axis}

        \begin{axis}[%
            hide axis,
            width=\figurewidthD,
            height=\figureheightD,
            xmin=0,
            xmax=64000,
            ymin=0,
            ymax=27,
            legend style={at={(1.125,2.1)},anchor=north east, draw=black,legend cell align=left, fill=white, row sep=0, inner sep=0, inner xsep=2pt,legend columns=6, font={\scriptsize}}
            ]

            \addlegendimage{thick, no marks, green}
            \addlegendentry{{\tt NKF}}
            \addlegendimage{thick, no marks, black}
            \addlegendentry{{\tt FDKF}}
            \addlegendimage{thick, no marks, blue}
            \addlegendentry{{\tt DLAC-Kalman}}
            \addlegendimage{thick, no marks, orange}
            \addlegendentry{{\tt NeuralKalman}}
            \addlegendimage{thick, no marks, magenta}
            \addlegendentry{{\tt DeepAdaptive}}
        \end{axis}
    
    \node (s1) at (-0.4,3){(dB)};
    \node (s1) at (14,-0.25){[s]};
    \end{tikzpicture}%
 
    \vspace{-7.5mm}
    \caption{Model performance averaged over the \textit{STFE sections} of all files with \textit{far-end speech excitation} in test set $\mathcal{D}_{\mathrm{test}}$. No nonlinearities are employed, but an RIR switch after 4s.} 
    \label{fig:over_time}
\end{figure}
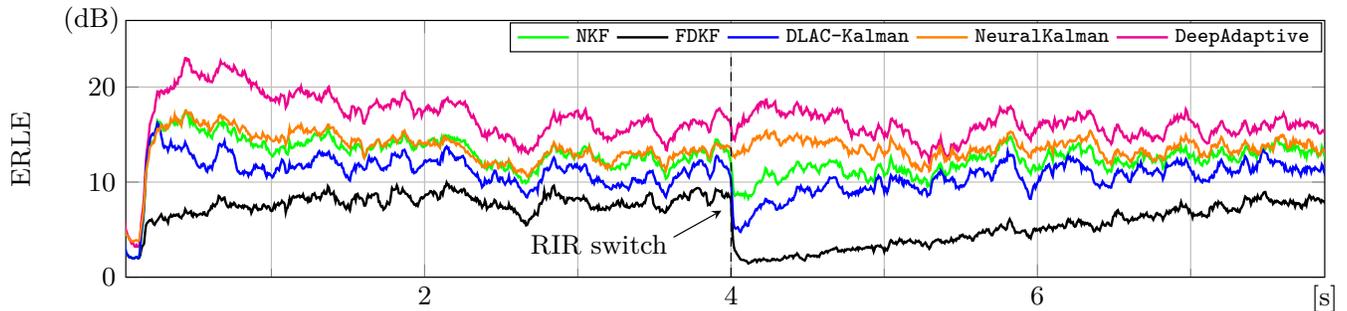

\begin{figure}[t]
    \centering
    	\small
    \newlength\figureheightC
    \newlength\figurewidthC
    \setlength\figureheightC{3cm}
    \setlength\figurewidthC{14cm}

    \definecolor{mycolor1}{rgb}{0.749019622802734,0,0.749019622802734}%
    \begin{tikzpicture}
        \begin{axis}[%
            width=\figurewidthC,
            height=\figureheightC,
            scale only axis,
            xmin=800,
            xmax=126000,
            xtick={0,32000,64000,96000,128000},
            xticklabels={0,2,4,6},
            xmajorgrids,
            xminorgrids,
            minor x tick num=1,
            scaled ticks=false,
            ymin=0,
            ymax=24,
            ytick={0,10,20},
            ylabel={ERLE},
            ymajorgrids,
            legend style={at={(0.993,0.0125)},anchor=south east,draw=black,legend cell align=left, fill=white, row sep=0, inner sep=0, inner xsep=2pt,legend columns=1, font={\scriptsize}}
            ]
		
                \addplot[ thick, no marks, black] file {fig/plots_20dB/WGN_STFE/Kalman_long8_eqD_CS_SER0_SNR0_BB_EoT.txt};
                
                \addplot[ thick, no marks, magenta] file {fig/plots_20dB/WGN_STFE/DeepAdaptive_AEC_model_8_LogMSE_LogMSEMS_ERNST_cluster_ern_CL_VCTKem_16_1080_1_full_fin_CS_SER0_SNR0_BB_EoT.txt};
                \addplot[ thick, no marks, green] file {fig/plots_20dB/WGN_STFE/NKF_AEC_model_8_LogMSE_LogMSEMS_ERNST_cluster_ern_CL_VCTKem_16_1080_1_full_stepped_CS_SER0_SNR0_BB_EoT.txt};

                \addplot[ thick, no marks, blue] file {fig/plots_20dB/WGN_STFE/DLAC_AEC_model_final_v2_noNorm_LogMSE_LogMSEMS_ERNST_cluster_ern_CL_VCTKem_16_1080_1_full_CS_SER0_SNR0_BB_EoT.txt};

                \addplot[ thick, no marks, orange] file {fig/plots_20dB/WGN_STFE/NeuralKalman_AEC_model_1_k_LogMSE_LogMSEMS_ERNST_cluster_ern_CL_VCTKem_16_1080_1_full_shortSeq_CS_SER0_SNR0_BB_EoT.txt};

                \draw [densely dashed, black] (axis cs: 64000,-30.0) -- (axis cs: 64000,45.0);

                \draw [black, ->,>={stealth}] (axis cs: 58000,5.2) -- (axis cs: 63000,8.0);

                \draw (axis cs: 52000,4.5) node (t) [text width=2cm] {RIR switch};

                \draw [black] (axis cs: 0,0.0) -- (axis cs: 640000,0.0);

        \end{axis}

        \begin{axis}[%
            hide axis,
            width=\figurewidthC,
            height=\figureheightC,
            xmin=0,
            xmax=64000,
            ymin=0,
            ymax=24,
            legend style={at={(1.125,2.1)},anchor=north east, draw=black,legend cell align=left, fill=white, row sep=0, inner sep=0, inner xsep=2pt,legend columns=6, font={\scriptsize}}
            ]

            \addlegendimage{thick, no marks, green}
            \addlegendentry{{\tt NKF}}
            \addlegendimage{thick, no marks, black}
            \addlegendentry{{\tt FDKF}}
            \addlegendimage{thick, no marks, blue}
            \addlegendentry{{\tt DLAC-Kalman}}
            \addlegendimage{thick, no marks, orange}
            \addlegendentry{{\tt NeuralKalman}}
            \addlegendimage{thick, no marks, magenta}
            \addlegendentry{{\tt DeepAdaptive}}
        \end{axis}
    
    \node (s1) at (-0.4,3){(dB)};
    \node (s1) at (14,-0.25){[s]};
    \end{tikzpicture}%
 
    \vspace{-7.5mm}
    \caption{Model performance averaged over the \textit{STFE sections} of all files with \textit{white Gaussian noise excitation} in test set $\mathcal{D}_{\mathrm{test}}^\mathrm{WGN}$. No nonlinearities are employed, but an RIR switch after 4s.} 
    \label{fig:over_time2}
\end{figure}
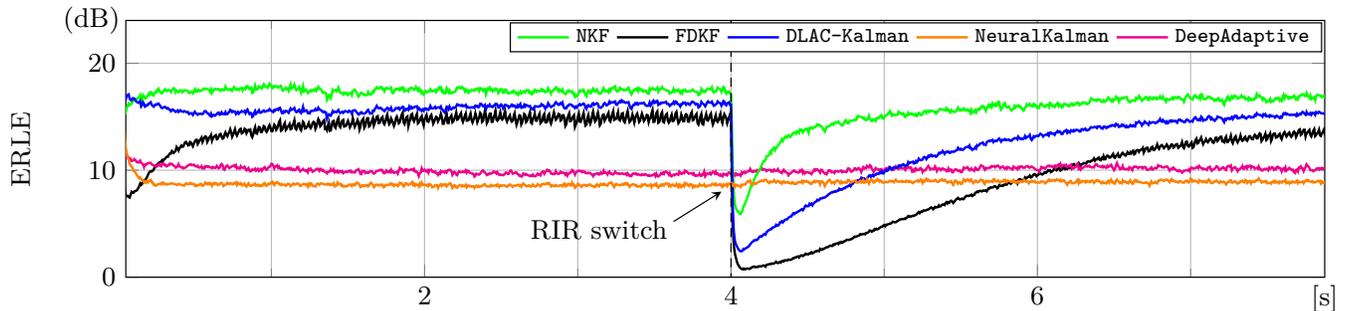

Fig.~\ref{fig:over_time2} reports on a similar case as Fig.~\ref{fig:over_time}, but now on the test set $\mathcal{D}_{\mathrm{test}}^{\mathrm{WGN}}$, which substitutes the speech FE excitation with white Gaussian noise (WGN). As before, no nonlinear distortions of the loudspeaker are present \mbox{($x'(n) = x(n)$)}. Note that WGN intentionally has \textit{not} been part of the training material. The WGN performance appears somewhat upside down in the sense that \texttt{NKF}, \texttt{DLAC-Kalman}, and \texttt{FDKF} reach a decent final accuracy, while depicting different rates of (re)convergence. This shows that also under conditions which are ideal for a classical adaptive filter (no nonlinearity, WGN excitation), hybrid models can outperform the classical \texttt{FDKF} in convergence and reconvergence speed. 
\texttt{DeepAdaptive} and \texttt{NeuralKalman} in this case are showcasing an unexpected limitation of ERLE to about $10$ dB only. 
The aforementioned masking behaviour confirms robustness to reconvergence also under WGN excitation (that was unseen in training). However, since the nonlinear distortion modules only see $\VEC{Y}_{\ell,k}$ and $\VEC{X}_{\ell,k}$ as inputs and the (noise-like) echo and background noise signals are much harder to separate than in previous test conditions, the models' reliance on the nonlinear distortion module as part of its mask estimation seems detrimental to the overall performance. This behaviour was found to be persistent for both models even when training with only WGN FE excitation.
For \texttt{DeepAdaptive} as the only investigated method computing a Kalman gain in total independence of an earlier AEC filter state $\widehat{H}_{\ell\!-\!1,k}$, we observed that the WGN cancellation is done in a very frequency-selective manner.

\subsection{Experiments: Speech Preservation vs.\ Echo Performance, Robustness Against Nonlinearity}
\label{ssec:NL}

\begin{figure}
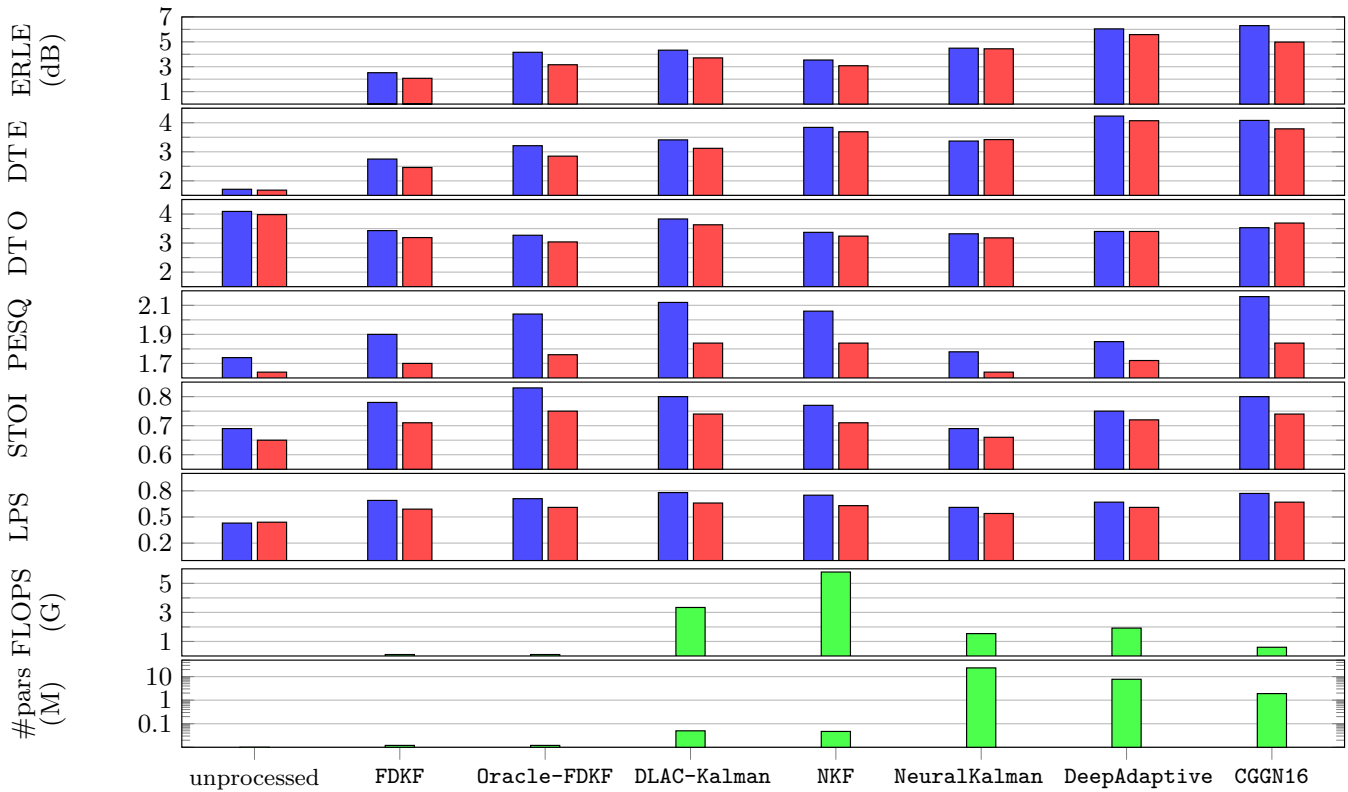

    \centering
    \includestandalone[width=1\columnwidth, mode=image|tex]{fig/fig7} \\
    \vspace{-2.0mm}
    \caption{Model performance averaged over the
     \textit{DT sections} of our test sets without nonlinear distortions ({\color{blue} $\mathcal{D}_{\mathrm{test}}$}, blue bars) and with nonlinear distortions ({\color{red} $\mathcal{D}_{\mathrm{test}}^\mathrm{NL}$}, red bars) of the loudspeaker signal. RIR switches take place in the middle of the DT sections. Metrics shown are the mean AECMOS (both DT Echo and DT Other),  PESQ, STOI, and LPS. The bottom panels display the number of giga floating point operations per second (FLOPS, green bars) and the number of parameters (\#pars, green bars, logarithmic scale!) related to each model.}
    \label{fig:means}
    \vspace{-3mm}
\end{figure}

While the previous ERLE-related experiments can provide a good understanding of the models' convergence behaviour, steady-state performance, and reconvergence behaviour over time, there is little information about the preservation of the NE speech component. For this reason, Fig.~\ref{fig:means} provides additional metrics evaluated on the DT sections of test set {$\mathcal{D}_{\mathrm{test}}$} (blue bars). Specifically, we provide DT Echo (DT E) as another metric for echo suppression effectiveness, DT Other (DT O) and PESQ to evaluate NE speech quality, and STOI to evaluate intelligibility of the NE speech signal. Furthermore, all metrics are also reported on the DT sections of test set $\mathcal{D}_{\mathrm{test}}^\mathrm{NL}$ (red bars), which features a harsh nonlinear distortion of the loudspeaker signal. RIR switches take place in the middle of the DT sections. We further added an {\tt Oracle-FDKF} experiment, where all instances of $E_{\ell,k}$ inside the ''Kalman Gain'' and ''Filter-State Update'' calculations have been replaced with $E^o_{\ell,k}=D_{\ell,k}-\widehat{D}_{\ell,k}$, meaning that the filter state is calculated undisturbed by other signal components. This acts as a reference of what the {\tt FDKF} could achieve under optimal circumstances. Furthermore, we included a convolutional gated GRU network with 16 layers ({\tt CGGN16}) \cite{Seidel2023}, which represents fully data-driven, masked-based AEC, as a model comparable to the previously observed masking behaviour of \texttt{NeuralKalman} and \texttt{DeepAdaptive}. The model was trained on the same dataset and target (AEC-only), with $K=512$ and $R=128$, introducing the same algorithmic delay as our neural Kalman filters.

First of all, we can see that the {\tt Oracle-FDKF} generally performs much better than the original {\tt FDKF}. However, DT\,O seems to not accurately report the actually improved performance of {\tt Oracle-FDKF} vs.\ {\tt FDKF} w.r.t.\ less residual echo and an equal or better near-end speech component. This aspect is further discussed on p.\ \pageref{subsec:Listening}. Interestingly, the oracle model is outperformed in almost every metric by neural Kalman methods. Even though {\tt Oracle-FDKF} has knowledge of the clean echo component, the conservative reconvergence speed of the classical calculation falls short of the DNN-augmented methods.

We observe that most of the scores under nonlinearity (red bars) are a bit behind the linear echo path case (blue bars). {\tt NeuralKalman} and \texttt{DeepAdaptive} with their nonlinear distortion modelling and masking behaviour are less impacted, even though their NE speech-related metrics remain low compared to the other neural Kalman methods. Generally, PESQ and STOI suffer considerably from a nonlinear echo path for all models. We conclude that residual nonlinear distortions can impact the NE speech signal quality significantly despite a smaller impact on an AEC metric such as ERLE.

Now we look at the blue bars only (linear echo case). The \texttt{DeepAdaptive} model delivers the highest level of echo cancellation (measured by DT Echo or ERLE) among the neural Kalman filters, which is a bit in disagreement with the example DT section in Fig.~\ref{fig:example}. This can be explained by the introduced RIR switch, which \texttt{DeepAdaptive} can handle easily as shown in Fig.~\ref{fig:over_time}.

In both the linear and nonlinear case, ERLE and DT Echo interestingly do not deliver the same rank orders. We attribute this observation to the fact that DT Echo reflects the subjective impression about echo removal \cite{Purin2021}, while ERLE simply measures the residual echo energy. AECMOS might overlook residual echo if it is no longer speech-like, as the underlying AECMOS neural network was trained on subjective listening test results, such that noise-like residual echo would be rated as less annoying or simply be not associated with the echo component anymore. ERLE, on the other hand, is penalized by each type of residual echo. 

When it comes to NE speech preservation, \texttt{DLAC-Kalman} achieves highest PESQ, STOI, and LPS scores among the neural Kalman filters, followed by \texttt{NKF} and \texttt{FDKF}. 
It seems that the \texttt{DLAC-Kalman} approach, which incorporates a DNN in the Kalman gain estimation alongside some remaining statistical components and traditional math, yields the best result at augmenting the \texttt{FDKF} without compromising NE speech quality. 
The good LPS scores also promise a good performance when using it as preprocessing, e.g., for ASR or keyword spotting.

Although \texttt{DLAC-Kalman}
and \texttt{NKF} employ small networks for estimating the Kalman gain, their good NE speech preservation is paid by a high computational complexity (bottom panels of Fig.\ \ref{fig:means}). This is simply because their Kalman gain DNN operates on isolated frequency bins. \texttt{NKF} additionally uses complex layers, which increases computational complexity even further. In contrast, \texttt{NeuralKalman} and \texttt{DeepAdaptive} mostly operate fully connected over all frequency bins. This raises the number of parameters significantly, but allows operation with less FLOPS. As we adjusted these models to be comparable in delay and effective reference input length, the displayed parameter counts differ from the numbers reported by the authors. Note that the masking behaviour of \texttt{NeuralKalman} and \texttt{DeepAdaptive} could give these models an advantage over the other neural Kalman filters. Often, subtractive AEC is augmented with a postfilter, e.g., to deal with nonlinearities and long echo paths. This is partly reflected in the superior echo suppression shown by {\tt DeepAdaptive}. Interestingly, when comparing to {\tt CGGN16}, we can see that apart from slightly lower ERLE on {$\mathcal{D}_{\mathrm{test}}^\mathrm{NL}$} and DT\,E, the end-to-end {\tt CGGN16} model clearly outperforms both \texttt{NeuralKalman} and \texttt{DeepAdaptive} as fully connected approaches while requiring less FLOPS and parameters.
For very low-power devices, expectedly, the \texttt{FDKF} is the unbeaten champion of resource efficiency, both with respect to parameters and with respect to computational complexity.

\subsection{Experiments: Informal Subjective Listening}
\label{subsec:Listening}
In this section, we report on our impressions from informal subjective listening to 106 files drawn from all of the aforementioned conditions (STFE with/without strong nonlinear distortion and with RIR switch, and DT with/without strong nonlinear distortion and with RIR switch). In general, both \texttt{NKF} and \texttt{DeepAdaptive} were among the subjectively top-ranking methods, however, depending on the underlying condition, one or the other was first. In particular, \texttt{DeepAdaptive} exhibited an aggressive echo cancellation both during double-talk and STFE (associated to highest ERLE and DT Echo scores in Fig.\ \ref{fig:means}), while \texttt{NKF} rendered the far-end echo to an inaudible enough level (in case of DT) and unintelligible enough to keep the echo annoyance perceptually at a minimum (in case of STFE). Quite regularly, \texttt{DeepAdaptive} showed an aggressive behavior especially by gating NE components at onsets of far-end signal activity, hence delivering a choppy voice because of distorting the onset/offset of NE (associated to its lower speech intelligibility shown in Fig.~\ref{fig:means}). 

\texttt{NKF} delivers the highest perceived  quality of the near-end speech component,
although lacking behind {\tt DLAC-Kalman} in DT\,O. This contradiction may partly stem from the fact that the subjective listening experiments leading to the DT\,O metric asked naive crowd workers ''How would you judge other degradation (noise, missing audio, distortions, cut-outs)?'' \cite{Purin2021}, which was potentially not clearly understood as excluding residual echo degradation, thereby weakening DT\,O's capability to quantify NE speaker distortion. This hypothesis is further underlined by our finding on pp. \pageref{ssec:NL}ff., that also {\tt Oracle-FDKF} performs inferior to {\tt FDKF} in DT\,O, which is definitely not supported from informal subjective listening and expert judgement of near-end speech component distortions.
We found the poor predictive performance of DT\,O particularly true in low SER conditions, which aligns with observations reported in \cite{Shachar2022}.

Back to \texttt{NKF}: While it delivers the highest perceived speech quality of the near-end component, it shows a very good trade-off versus the far-end echo removal, often equal or even better than the \texttt{DeepAdaptive} approach (e.g., during double-talk or after a RIR switch). In case of a male maskee (far-end) and female masker (near-end), however, \texttt{DeepAdaptive} was better in terms of both residual echo suppression and consequently improving both the intelligibility and naturalness of the near-end signal during double-talk. It was also observed that in some rare instances, {\tt NKF} displays instability issues during DT, leading to an overall slightly lower than expected scoring. For \texttt{NeuralKalman}, the OLS algorithm seemed to cause artifacts during convergence, similar to the findings in \cite{Casebeer2024}. The remaining methods, i.e., \texttt{DLAC-Kalman} and \texttt{FDKF}, 
{appeared with clearly insufficient echo cancellation in our subjective listening, especially in case of the RIR switch and nonlinearity, confirming their traditional dependency on system extension with postfiltering for residual echo suppression (cf.\ Fig.\ \ref{fig:systemlvl}). Their near-end speech component, however, was well preserved.}

\subsection{Consequences and Recommendations for Hybrid System Design}

Based on our findings, we would like to draw a few conclusions on hybrid system design when facing a limited computational complexity, memory, latency, or data budget. As discussed on pp.\ \pageref{sec:Hybrid}ff.\ and pp.\ \pageref{ssec:NL}ff., the investigated models exhibit different trade-offs regarding computational complexity and parameter footprint. Per-bin processing ({\tt NKF}, {\tt DLAC-Kalman}) allows for low parameter counts, and there might be an efficient way to parallelize these structures to compensate for the overall high computational demand. Furthermore, their architecture allows a decoupling of frame rate and update rate: To save on computational complexity, e.g., one could calculate the filter update every 2nd or 4th frame. Due to their per-bin processing, it could also be feasible to employ a single trained model at variable filter lengths, allowing adjustment for different environments and latency demands. Generally, the amount of required training data is proportional to model size, which favors such models as well. When it comes to fully connected neural Kalman filters ({\tt DeepAdaptive}, {\tt NeuralKalman}) their advantage lies in lower computational complexity, but they scale less well with higher filter lengths and provide no flexibility, especially with the distortion modelling approach preventing decoupling of frame rate and update rate. Furthermore, if one decides against a per-bin hybrid AEC, among the non-flexible AEC approaches, the end-to-end solution (e.g., {\tt CGGN16}) turns out to be highly preferable overall, both in complexity and performance. When it comes to the choice between OLA and OLS, the OLS algorithm in hybrid AEC seems to cause artifacts during rapid convergence, and we generally found it more prone to instabilities during training.

It should also be mentioned that, while the neural Kalman filters are improving performance over the classical FDKF, the observed results are by no means perfect. Some residual echo remains due to nonlinearities and long echo tails. To that extent, a postfilter, as commonly employed with classical AEC solutions, can be beneficial. This is especially true for {\tt NKF} and {\tt DLAC-Kalman}, while the discussed masking behaviour of {\tt NeuralKalman} and {\tt DeepAdaptive} might already comprise some of that functionality.  

\section{Conclusions}
\label{sec:Conclusions}

{This article initially revisited the genesis and formal background of the original frequency-domain adaptive Kalman filter (FDKF) as a baseline and framework for many extensions recently proposed with elements from neural networks. Under this framework and with common mathematical notation, we made explicit where exactly neural networks replace parts of the FDKF. The structural analysis and experimental study has taken place in the regime of the  acoustic echo cancellation problem for hands-free systems and used a consistent simulation framework and data to draw relevant conclusions for future development of neural Kalman filters. Specifically, it was shown that neural Kalman filters reveal better echo performance than FDKF along with better (re)convergence. While FDKF has been traditionally very strong in preserving single-talk near-end speech,
some neural Kalman filters depict potential to preserve NE speech even better than the classical approach during double talk.} 
Especially approaches utilizing per-bin processing provide an attractive solution due to their flexibility and small parameter count, and in case of limited training data availability.
Fully connected neural Kalman filters come with none of these advantages, and are outperformed by an end-to-end AEC in most metrics and resource requirements.

\section{Acknowledgment}
\label{sec:Ack}

We would like to thank Thomas Haubner and Yixuan Zhang for providing details and assistance in setting up their respective \texttt{DLAC-Kalman} and \texttt{NeuralKalman} models. We acknowledge the \texttt{NKF} authors for providing code on GitHub. 
Additionally, we would like to thank the anonymous reviewers of this article for their insightful observations and valuable contributions to model discussions.

\newpage

\section*{References}
\begin{multicols}{2}
\renewcommand{\refname}{ \vspace{-\baselineskip}\vspace{-1.6mm} }
\bibliographystyle{ieeetr}
\bibliography{IEEEabrv,MGCabrv,strings,NKF,refs}

\begin{thebibliography}{10}

\bibitem{Haensler2004}
E.~Hänsler and G.~Schmidt, {\em Acoustic Echo and Noise Control: A Practical Approach}.
\newblock Hoboken, NJ, USA: John Wiley \& Sons, Ltd, 2004.

\bibitem{Kellermann88}
W.~Kellermann, ``{Analysis and Design of Multirate Systems for Cancellation of Acoustical Echoes},'' in {\em Proc.\ of ICASSP}, (New York, NY, USA), pp.~2570--2573, Apr. 1988.

\bibitem{Breining99}
C.~Breining, P.~Dreiseitel, E.~H{\"a}nsler, A.~Mader, B.~Nitsch, H.~Puder, T.~Schertler, G.~Schmidt, and J.~Tilp, ``{Acoustic Echo Control: An Application of Very-High-Order Adaptive Filters},'' {\em {IEEE} Signal Process. Mag.}, vol.~16, pp.~42--69, July 1999.

\bibitem{Benesty2001}
J.~Benesty, T.~G{\"a}nsler, D.~Morgan, M.~Sondhi, and S.~Gay, {\em {Advances in Network and Acoustic Echo Cancellation}}.
\newblock Berlin, Germany: Springer, 2001.

\bibitem{Enzner2006}
G.~Enzner and P.~Vary, ``{Frequency-Domain Adaptive Kalman Filter for Acoustic Echo Control in Hands-Free Telephones},'' {\em Signal Processing}, vol.~86, pp.~1140--1156, June 2006.

\bibitem{Richard2023}
G.~Richard, P.~Smaragdis, S.~Gannot, P.~A. Naylor, S.~Makino, W.~Kellermann, and A.~Sugiyama, ``{Audio Signal Processing in the 21st Century: The Important Outcomes of the Past 25 Years},'' {\em {IEEE} Signal Process. Mag.}, vol.~40, no.~5, pp.~12--26, 2023.

\bibitem{Franzen2021}
J.~Franzen, E.~Seidel, and T.~Fingscheidt, ``{AEC in A Netshell: On Target and Topology Choices for FCRN Acoustic Echo Cancellation},'' in {\em Proc.\ of ICASSP}, (Toronto, ON, Canada), pp.~156--160, June 2021.

\bibitem{Seidel2021}
E.~Seidel, J.~Franzen, M.~Strake, and T.~Fingscheidt, ``{Y$^2$-Net FCRN for Acoustic Echo and Noise Suppression},'' in {\em Proc. of Interspeech}, (Brno, Czech Republic), pp.~4763--4767, Oct. 2021.

\bibitem{Braun2022}
S.~Braun and M.~L. Valero, ``{Task Splitting for DNN-Based Acoustic Echo and Noise Removal},'' in {\em Proc.\ of IWAENC}, (Bamberg, Germany), pp.~386--390, Sept. 2022.

\bibitem{Seidel2023}
E.~Seidel, P.~Mowlaee, and T.~Fingscheidt, ``{Efficient Deep Acoustic Echo Suppression with Condition-Aware Training},'' in {\em Proc. of WASPAA}, (New Paltz, NY, USA), pp.~1--5, Oct. 2023.

\bibitem{Wang2019_entryGE}
H.~Zhang, K.~Tan, and D.~Wang, ``{Deep Learning for Joint Acoustic Echo and Noise Cancellation with Nonlinear Distortions},'' in {\em Proc. of Interspeech}, (Graz, Austria), pp.~4255--4259, Sept. 2019.

\bibitem{Valin_AEC}
J.-M. Valin, S.~Tenneti, K.~Helwani, U.~Isik, and A.~Krishnaswamy, ``{Low-Complexity, Real-Time Joint Neural Echo Control and Speech Enhancement Based On PercepNet},'' in {\em Proc.\ of ICASSP}, (Toronto, ON, Canada), pp.~7133--7137, June 2021.

\bibitem{halimeh_combining_2021}
M.~Halimeh, T.~Haubner, A.~Briegleb, A.~Schmidt, and W.~Kellermann, ``{Combining Adaptive Filtering and Complex-Valued Deep Postfiltering for Acoustic Echo Cancellation},'' in {\em Proc.\ of ICASSP}, (Toronto, ON, Canada), pp.~121--125, June 2021.

\bibitem{ivry_deep_2021}
A.~Ivry, I.~Cohen, and B.~Berdugo, ``{Deep Residual Echo Suppression with a Tunable Tradeoff Between Signal Distortion and Echo Suppression},'' in {\em Proc.\ of ICASSP}, (Toronto, ON, Canada), pp.~126--130, June 2021.

\bibitem{Revach2022}
G.~Revach, N.~Shlezinger, X.~Ni, A.~L. Escoriza, R.~J.~G. van Sloun, and Y.~C. Eldar, ``{KalmanNet: Neural Network Aided Kalman Filtering for Partially Known Dynamics},'' {\em IEEE Trans. Sig. Proc.}, vol.~70, p.~1532–1547, Jan. 2022.

\bibitem{MetaAF2023}
J.~Casebeer, N.~J. Bryan, and P.~Smaragdis, ``{Meta-AF: Meta-Learning for Adaptive Filters},'' {\em {IEEE} T-ASLP}, vol.~31, pp.~355--370, 2023.

\bibitem{Enzner2010}
G.~Enzner, ``{Bayesian Inference Model for Applications of Time-Varying Acoustic System Identification},'' in {\em Proc.\ of EUSIPCO}, pp.~2126--2130, Aug. 2010.

\bibitem{Paleologu2013}
C.~Paleologu, J.~Benesty, and S.~Ciochina, ``{Study of the General Kalman Filter for Echo Cancellation},'' {\em {IEEE} T-ASLP}, vol.~21, pp.~1539--1549, Aug. 2013.

\bibitem{Jung2014}
M.-A. Jung, S.~Elshamy, and T.~Fing\-scheidt, ``{An Automotive Wideband Stereo Acoustic Echo Canceler Using Frequency-Domain Adaptive Filtering},'' in {\em Proc.\ of EUSIPCO}, (Lisbon, Portugal), pp.~1452--1456, Sept. 2014.

\bibitem{yang2017frequency}
F.~Yang, G.~Enzner, and J.~Yang, ``{Frequency-Domain Adaptive {Kalman} Filter with Fast Recovery of Abrupt Echo-Path Changes},'' {\em {IEEE} SP Letters}, vol.~24, pp.~1778--1782, June 2017.

\bibitem{Schrammen2019}
M.~Schrammen, S.~Kühl, S.~Markovich-Golan, and P.~Jax, ``{Efficient Nonlinear Acoustic Echo Cancellation by Dual-Stage Multi-Channel {Kalman} Filtering},'' in {\em Proc.\ of ICASSP}, (Brighton, UK), pp.~975--979, May 2019.

\bibitem{Cutler2023}
R.~Cutler, A.~Saabas, T.~Parnamaa, M.~Purin, E.~Indenbom, N.-C. Ristea, J.~Guzhvin, H.~Gamper, S.~Braun, and R.~Aichner, ``{ICASSP 2023 Acoustic Echo Cancellation Challenge},'' {\em arXiv preprint:2309.12553}, Sept. 2023.

\bibitem{Franzen2018a}
J.~Franzen and T.~Fing\-scheidt, ``{An Efficient Residual Echo Supression for Multi-Channel Acoustic Echo Cancellation Based on the Frequency-Domain Adaptive Kalman Filter},'' in {\em Proc.\ of ICASSP}, (Calgary, Canada), pp.~226--230, Apr. 2018.

\bibitem{Seidel2024}
E.~Seidel, P.~Mowlaee, and T.~Fingscheidt, ``{Convergence and Performance Analysis of Classical, Hybrid, and Deep Acoustic Echo Control},'' {\em IEEE T-ASLP}, vol.~32, pp.~2857--2870, 2024.

\bibitem{Haykin2002}
S.~Haykin, {\em {Adaptive Filter Theory}}.
\newblock Hoboken, NJ, USA: Prentice-Hall, 2002.

\bibitem{Gay2000}
S.~L. Gay and J.~Benesty, eds., {\em Acoustic Signal Processing for Telecommunication}.
\newblock USA: Kluwer Academic Publishers, 2000.

\bibitem{Vary2006}
P.~Vary and R.~Martin, {\em Digital Speech Transmission}.
\newblock Hoboken, NJ, USA: John Wiley \& Sons, Ltd, 2006.

\bibitem{Scharf91}
L.~L. Scharf, {\em Statistical Signal Processing}.
\newblock Addison-Wesley Publishing Company, 1991.

\bibitem{Malik2011}
S.~Malik and G.~Enzner, ``{Recursive Bayesian Control of Multichannel Acoustic Echo Cancellation},'' {\em {IEEE} SP Letters}, vol.~18, pp.~619--622, Nov. 2011.

\bibitem{Ferrara80}
E.~Ferrara, ``{Fast Implementation of {LMS} Adaptive Filters},'' vol.~28, pp.~474--475, August 1980.

\bibitem{Kuech2014}
F.~Kuech, E.~Mabande, and G.~Enzner, ``{State-Space Architecture of the Partitioned-Block-Based Acoustic Echo Controller},'' in {\em Proc.\ of ICASSP}, (Florence, Italy), pp.~1295--1299, May 2014.

\bibitem{Haubner2024}
T.~Haubner, A.~Brendel, and W.~Kellermann, ``{End-to-End Deep Learning-Based Adaptation Control for Linear Acoustic Echo Cancellation},'' {\em {IEEE T-ASLP}}, vol.~32, pp.~227--238, 2024.

\bibitem{Yang2023}
D.~Yang, F.~Jiang, W.~Wu, X.~Fang, and M.~Cao, ``{Low-Complexity Acoustic Echo Cancellation with Neural Kalman Filtering},'' in {\em Proc.\ of ICASSP}, (Rhodes Island, Greece), pp.~7846--7850, June 2023.

\bibitem{Zhang2023}
Y.~Zhang, M.~Yu, H.~Zhang, D.~Yu, and D.~Wang, ``{NeuralKal\-man: A Learnable Kalman Filter for Acoustic Echo Cancellation},'' {\em arXiv preprint:2301.12363}, Jan. 2023.

\bibitem{Zhang2022d}
H.~Zhang, S.~Kandadai, H.~Rao, M.~Kim, T.~Pruthi, and T.~Kristjansson, ``{Deep Adaptive AEC: Hybrid of Deep Learning and Adaptive Acoustic Echo Cancellation},'' in {\em Proc.\ of ICASSP}, (Singapore), pp.~756--760, May 2022.

\bibitem{Mack2020}
W.~Mack and E.~A.~P. Habets, ``{Deep Filtering: Signal Extraction and Reconstruction Using Complex Time-Frequency Filters},'' {\em IEEE Signal Processing Letters}, vol.~27, pp.~61--65, 2020.

\bibitem{Purin2021}
M.~Purin, S.~Sootla, M.~Sponza, A.~Saabas, and R.~Cutler, ``{AECMOS: A Speech Quality Assessment Metric for Echo Impairment},'' in {\em Proc.\ of ICASSP}, (Singapore), pp.~901--905, May 2022.

\bibitem{ITU_P862.2_Corr2}
{\em {ITU-T Rec. P.862.2 Corrigendum 1: Wideband Extension to Rec. P.862 for the Assessment of Wideband Telephone Networks and Speech Codecs}}, Oct. 2017.

\bibitem{Taal2011}
C.~H. Taal, R.~C. Hendriks, R.~Heusdens, and J.~Jensen, ``{An Algorithm for Intelligibility Prediction of Time–Frequency Weighted Noisy Speech},'' {\em {IEEE} T-ASLP}, vol.~19, no.~7, pp.~2125--2136, 2011.

\bibitem{Paszke2019}
A.~Paszke, S.~Gross, F.~Massa, A.~Lerer, J.~Bradbury, G.~Chanan, T.~Killeen, Z.~Lin, N.~Gimelshein, L.~Antiga, A.~Desmaison, {\em et~al.}, ``{PyTorch: An Imperative Style, High-Performance Deep Learn\-ing Library},'' in {\em Proc. of NeurIPS}, (Vancouver, BC, Canada), pp.~8024--8035, Dec. 2019.

\bibitem{VCTK}
J.~Yamagishi, C.~Veaux, and K.~MacDonald, ``{CSTR VCTK Corpus: Eng\-lish Multi-Speaker Corpus for CSTR Voice Cloning Toolkit}.'' {University of Edinburgh. The Centre for Speech Technology Research}, 2017.

\bibitem{Klippel2006}
W.~Klippel, ``{Tutorial: Loudspeaker Nonlinearities -- Causes, Parameters, Symptoms},'' {\em Journal of the Audio Engineering Society}, vol.~54, pp.~907--939, Oct. 2006.

\bibitem{Thiemann2013}
J.~Thiemann, N.~Ito, and E.~Vincent, ``{The Diverse Environments Multi-Channel Acoustic Noise Database: A Database of Multichannel Environmental Noise Recordings},'' {\em The Journal of the Acoustical Society of America}, vol.~133, no.~5, pp.~3591--3591, 2013.

\bibitem{QUT}
D.~B. Dean, S.~Sridharan, R.~J. Vogt, and M.~W. Mason, ``{The QUT-NOISE-TIMIT Corpus for the Evaluation of Voice Activity Detection Algorithms},'' in {\em Proc. of Interspeech}, (Makuhari, Japan), p.~3110–3113, Sept. 2010.

\bibitem{Kingma2015}
D.~P. Kingma and J.~Ba, ``{Adam: A Method for Stochastic Optimization},'' in {\em Proc. of ICLR}, (San Diego, CA, USA), pp.~1--15, May 2015.

\bibitem{TIMIT}
J.~S. Garofolo, L.~F. Lamel, W.~M. Fisher, J.~G. Fiscus, and D.~S. Pallett, ``{TIMIT Acoustic-Phonetic Continous Speech Corpus}.'' {Linguistic Data Consortium, Philadelphia, PA, USA}, 1993.

\bibitem{Jeub2010}
M.~Jeub, M.~Schäfer, H.~Krüger, C.~M. Nelke, C.~Beaugeant, and P.~Vary, ``{Do We Need Dereverberation for Hand-Held Telephony?},'' in {\em Proc. of ICA}, (Sydney, Australia), pp.~3793--3799, Aug. 2010.

\bibitem{ETSI2008}
ETSI EG 202 396-1, {\em {Speech Processing, Transmission and Quality Aspects (STQ); Speech Quality Performance in the Presence of Background Noise; Part 1: Background Noise Simulation Technique and Background Noise Database}}, Sept. 2008.

\bibitem{Zhang18g}
H.~Zhang and D.~Wang, ``{Deep Learning for Acoustic Echo Cancellation in Noisy and Double-Talk Scenarios},'' in {\em Proc. of Interspeech}, (Hyderabad, India), pp.~3239--3243, Sept. 2018.

\bibitem{Pirklbauer2023}
J.~Pirklbauer, M.~Sach, K.~Fluyt, W.~Tirry, W.~Wardah, S.~Moeller, and T.~Fingscheidt, ``{Evaluation Metrics for Generative Speech Enhancement Methods: Issues and Perspectives},'' in {\em Proc.\ of 15th ITG Conference on Speech Communication}, (Aachen, Germany), pp.~265--269, Sep 2023.

\bibitem{Shachar2022}
E.~Shachar, I.~Cohen, and B.~Berdugo, ``{Double-Talk Detection-Aided Residual Echo Suppression via Spectrogram Masking and Refinement},'' {\em Acoustics}, vol.~4, no.~3, pp.~637--655, 2022.

\bibitem{Casebeer2024}
J.~Casebeer, N.~J. Bryan, and P.~Smaragdis, ``{Scaling Up Adaptive Filter Optimizers},'' {\em arXiv preprint:2403.00977}, Mar. 2024.

\end{thebibliography}
\end{multicols}

\section*{Biographies}
\label{sec:bio}

\begin{IEEEbiographynophoto}{Ernst Seidel} (e.seidel@tu-bs.de) received the M.Sc.\ degree in electrical engineering in 2021 from Techni\-sche Universität Braunschweig, Germany, and is currently enrolled in his Ph.D.\ studies at the Institute for Communications Technology, Technische Universität Braunschweig, in the field of machine-learned acoustic echo cancellation. As part of his research, he designs and evaluates classical, machine-learned, and hybrid approaches to acoustic echo cancellation and suppression. His work achieved the 4th rank in the Interspeech 2021 Acoustic Echo Cancellation Challenge. Seidel is an IEEE Member.
\end{IEEEbiographynophoto}

\begin{IEEEbiographynophoto}{Gerald Enzner} (gerald.enzner@uol.de) received his Dr.-Ing.\ degree  from RWTH Aachen University (2006). He was with Ruhr-University Bochum (2007-2021) and was there appointed to faculty via Habilitation. Since 2021, he joined Carl-von-Ossietzky University Oldenburg as a Professor of Speech Technology and Hearing Aids. His research interests include signal processing, adaptive filtering, and machine learning in problems of speech enhancement and binaural modeling. He has been a member of ITG's Speech Acoustics Committee since 2012, of IEEE's Audio and Acoustic Signal Processing Technical Committee since 2020, and Chair of IEEE's Signal Processing Society Germany Chapter since 2024. He is IEEE Senior Member.\end{IEEEbiographynophoto}

\begin{IEEEbiographynophoto}{Pejman Mowlaee} (pmowlaee@jabra.com) received the Ph.D.\ degree from Aalborg University, Denmark (2010). From 2011-2012 he was a Postdoc fellow with Marie Curie (AUDIS) at the Institute of Communication Acoustics at Ruhr-University Bochum, Germany. From 2012 to 2017, he was an Assistant Professor at Graz University of Technology, Austria, and in 2017 appointed to faculty via Habilitation. He was with WS Audiology between 2017-2020 and since 2020 with GN Advanced Science Denmark. He is an Associate Editor for the IEEE T-ASLP, member of the IEEE Signal Processing Society Technical Committee on Audio and Acoustic Signal Processing, and IEEE Senior Member.
\end{IEEEbiographynophoto}

\begin{IEEEbiographynophoto}
{Tim Fingscheidt} (t.fingscheidt@tu-bs.de)  received his Ph.D.\ degree in 1998 from RWTH Aachen University. He was with AT\&T Labs, USA, and Siemens, Munich. Since 2006, he has been a Full Professor with TU Braunschweig, Germany. His research interests include signal processing, machine learning, speech processing and computer vision. He received several awards, among them an ITG Award and three CVPR-W Best Paper Awards. He served as Associate Editor for the IEEE T-ASLP (2008–2010), was member of the IEEE Speech and Language Processing Technical Committee (2012–2014 and 2016–2018). He is ITG Fellow, ITG Board Member, and IEEE Senior Member.
\end{IEEEbiographynophoto}

\vfill

\end{document}


\title{\!Neural Kalman Filters for Acoustic Echo Cancellation\!

{\large {White paper "IEEE SPM Special Issue on Model-based and Data-Driven Audio Signal Processing"}}}

\author{Ernst Seidel$^{\ast}$, Gerald Enzner$^{\circ}$, Pejman Mowlaee$^{\vardiamond}$, Tim Fingscheidt$^{\ast}$ \\
{\small e.seidel@tu-bs.de, gerald.enzner@uni-oldenburg.de, pmowlaee@jabra.com, t.fingscheidt@tu-braunschweig.de} \\[0.3cm]


$^{\ast}$Institute for Communications Technology,
Technische Universität Braunschweig, Germany\\ 
$^{\circ}$Division of Speech Technology and Hearing Aids, Carl von Ossietzky University Oldenburg, Germany\\
$^{\vardiamond}$GN Audio A/S, Ballerup, Denmark
}

\maketitle

\vspace{-7ex}
\section{Short bio for each author}
\par {\bf Tim Fingscheidt} (Senior Member, IEEE) received the Dipl.-Ing. degree in electrical engineering in 1993 and the Ph.D.\ degree in 1998 from RWTH Aachen University, Germany. He joined AT\&T Labs, Florham Park, NJ, USA, in 1998, and Siemens AG (Mobile Devices), Munich, Germany, in 1999, being responsible for speech technology R\&D in Siemens mobile phones worldwide. He then was with Siemens Corporate Technology, Munich, Germany (2005–2006). Since 2006, he has been a Full Professor with the Institute for Communications Technology, Technische Universität Braunschweig, Germany. His research interests include signal processing and machine learning, with applications in speech processing and computer vision. He is the recipient of several awards, including the Vodafone Mobile Communications Foundation prize in 1999 and the 2002 ITG Award of the Association of German Electrical Engineers (VDE ITG). In 2014, he received a 10,000 Euro technology transfer prize for probably the first large-scale application of FDKF for acoustic echo cancellation in high-end enterprise phones. In 2017 and 2020, he co-authored an ITG award-winning publication, and in 2019, 2020, and 2021 he was given the Best Paper Award of a CVPR Workshop. He has been the Speaker of the Speech Acoustics Committee ITG AT3 since 2015. He was an Associate Editor for the IEEE Transactions on Audio, Speech, and Language Processing (2008–2010), and for the EURASIP Journal on Audio, Speech, and Music Processing (2013-2018). He served as a member of the IEEE Speech and Language Processing Technical Committee (2012–2014, 2016–2018).

\par {\bf Gerald Enzner} (Senior Member, IEEE) received the Dipl.-Ing.\ degree from University of Erlangen-Nuremberg, Germany, 2000, and the Dr.-Ing.\ degree from RWTH Aachen University, Germany, 2006, both in electrical engineering. Since 2007, he has been a Principal Investigator with the Institute of Communication Acoustics at Ruhr-University Bochum, Germany, where the FDKF theme frequently was in the core of his research programs for acoustic signal enhancement. 
For a publication related to the FDKF, he received a 
Best Paper Award on IWAENC 2012.
From 2009 to 2011, he was a parallel member of the Global Young Faculty of the University Alliance Metropolis Ruhr and the Mercator Foundation. He then was appointed to faculty via Habilitation (2013) and Inauguration (2016) in the Department of Electrical Engineering and Information Technology at Ruhr-University Bochum. In 2021, he joined the Department of Medical Physics and Acoustics at Carl-von-Ossietzky University Oldenburg as a Professor of speech technology and hearing aids. His research interests include the methods of statistical signal processing, adaptive filtering, and machine learning for problems of speech enhancement and binaural modeling. He has been member of VDE/ITG's Speech Acoustics Committee since 2012, of IEEE's Audio and Acoustic Signal Processing Technical Committee since 2020, and Vice-Chair of IEEE's Signal Processing Society Germany Chapter since 2016.

\par \textbf{Pejman Mowlaee} (Senior Member, IEEE) received the M.Sc. degree in Communication Systems from Iran University of Science \& Technology, Iran, in 2006 and Ph.D.\ degree in Signal Processing from Aalborg Unviersity, Denmark in 2010. From 2011 to 2012 he was a Postdoc fellow with Marie Curie Programme AUDIS (Digital Signal Processing in Audiology) at the Institute of Communication Acoustics, Ruhr Universit\"{a}t Bochum, Germany. From 2012 to 2017, he was an Assistant Professor at Graz University of Technology, Austria and in 2017 appointed as Privatdozent with Habilitation in speech signal processing and Adjunct Professor. Between 2017-2020 he was a signal processing specialist with WS Audiology A/S, Denmark. Since 2020 he is a lead research scientist at GN Audio A/S with main responsibility in scouting and development of new signal processing concepts for speech communication devices. He is the co-author of the book “Single Channel Phase-Aware Signal Processing in Speech Communication: Theory and Practice” John Wiley \& Sons 2017. His research interests include signal processing and machine learning for audio applications. He is an Associate Editor for the IEEE/ACM Transactions on Audio, Speech and Language Processing and an elected Member of the IEEE Signal Processing Society Technical Committee on Audio and Acoustic Signal Processing.

\par {\bf Ernst Seidel} received the M.Sc. degree in electrical engineering in 2021 from Technische Universität Braunschweig, Germany and is currently enrolled in his Ph.D.\ studies at the Institute for Communications Technology, Technische Universität Braunschweig, in the field of machine-learned acoustic echo cancellation. As part of his research, he designs and evaluates classical, machine-learned, and hybrid approaches to acoustic echo cancellation and suppression. His work achieved the 4th place in the Interspeech 2021 Acoustic Echo Cancellation Challenge.
